\pdfoutput=1
\documentclass[sigconf,nonacm]{acmart}

\usepackage{booktabs}
\usepackage{array}
\usepackage{listings}
\usepackage{xspace}
\usepackage{tikz}
\usetikzlibrary{positioning,arrows.meta}
\usepackage{pgfplots}
\pgfplotsset{compat=1.16}
\pgfplotsset{every axis/.append style={
  ymajorgrids, major grid style={draw=black!12, line width=0.3pt},
  tick align=outside, tick pos=left,
  axis line style={black!80}, every tick/.style={black!80}}}
\usepackage{subcaption}
\usepackage{algorithm}
\usepackage{algpseudocode}

\definecolor{tsOurs}{HTML}{C0643F}
\definecolor{tsOursLight}{HTML}{F2C19C}
\definecolor{tsFull}{HTML}{3D4852}
\definecolor{tsBase}{HTML}{4C7090}
\definecolor{tsBaseMid}{HTML}{8DAAC2}
\definecolor{tsBaseLight}{HTML}{CCDAE5}

\let\origunderscore\_
\renewcommand{\_}{\origunderscore\allowbreak}
\newcommand{\sys}{TapeSAC\xspace}

\newcommand{\closure}{\mathrm{closure}}

\begin{document}

\title{TapeSAC: Automated, Pareto-Optimal Fine-Grained Memory Management for Large Model Training}

\settopmatter{authorsperrow=4}
\makeatletter
\renewcommand{\@authorfont}{\large}
\renewcommand{\@affiliationfont}{\small}
\makeatother

\author{Kaihui Ren}
\authornote{This work was conducted during the author's internship at KlingAI.}
\email{renkaihui@kuaishou.com}
\affiliation{%
  \institution{KlingAI Research}
  \city{Beijing}
  \country{China}
}

\author{Yifan Zhao}
\authornote{Project leader.}
\email{zhaoyifan14@kuaishou.com}
\affiliation{%
  \institution{KlingAI Research}
  \city{Shanghai}
  \country{China}
}

\author{Zejian Wang}
\authornotemark[1]
\email{wangzejian@kuaishou.com}
\affiliation{%
  \institution{KlingAI Research}
  \city{Shanghai}
  \country{China}
}

\author{Yingtong Xiong}
\email{xiongyingtong@kuaishou.com}
\affiliation{%
  \institution{KlingAI Research}
  \city{Shanghai}
  \country{China}
}

\author{Xiaoling Zhou}
\authornotemark[1]
\email{zhouxiaoling03@kuaishou.com}
\affiliation{%
  \institution{KlingAI Research}
  \city{Beijing}
  \country{China}
}

\author{Daohong Xu}
\email{xudaohong@kuaishou.com}
\affiliation{%
  \institution{KlingAI Research}
  \city{Beijing}
  \country{China}
}

\author{Chenhao Guan}
\authornotemark[1]
\email{guanchenhao@kuaishou.com}
\affiliation{%
  \institution{KlingAI Research}
  \city{Shanghai}
  \country{China}
}

\author{Zhikang Mo}
\email{mozhikang@kuaishou.com}
\affiliation{%
  \institution{KlingAI Research}
  \city{Shanghai}
  \country{China}
}

\author{Jihu Guo}
\authornotemark[1]
\email{guojihu@kuaishou.com}
\affiliation{%
  \institution{KlingAI Research}
  \city{Shanghai}
  \country{China}
}

\author{Zhuoru Ma}
\email{mazhuoru@kuaishou.com}
\affiliation{%
  \institution{KlingAI Research}
  \city{Shanghai}
  \country{China}
}

\author{Yufeng Li}
\email{liyufeng06@kuaishou.com}
\affiliation{%
  \institution{KlingAI Research}
  \city{Shanghai}
  \country{China}
}

\author{Lin Zhang}
\email{zhanglin22@kuaishou.com}
\affiliation{%
  \institution{KlingAI Research}
  \city{Shanghai}
  \country{China}
}

\author{Yiqiao Liao}
\authornote{Corresponding authors.}
\email{liaoyiqiao@kling.ai}
\affiliation{%
  \institution{KlingAI Research}
  \city{Beijing}
  \country{China}
}

\author{Chengru Song}
\authornotemark[3]
\email{songchengru@kuaishou.com}
\affiliation{%
  \institution{KlingAI Research}
  \city{Beijing}
  \country{China}
}

\renewcommand{\shortauthors}{Kling Infra Team}

\begin{abstract}
Selective activation checkpointing (SAC) decides which activations to
save for the backward pass and which to recompute: each choice is a
point in the plane of peak memory and recomputation time, and the goal is that plane's Pareto
frontier. Current practice falls short of this goal. Hand-tuned policies
fix a single point that does not adapt when the model, batch or kernels
change; PyTorch's SAC reruns each block's forward pass in program order,
recomputing values the backward pass never reads; and the memory budget of
\texttt{torch.compile} treats recomputation as free or forbidden and
selects saved tensors by bytes alone, an assumption that fails for the
custom kernels that produce many of a large model's activations.

We present \sys, which automates SAC in eager PyTorch without editing
the model's forward code. \sys records each block's forward pass at the dispatcher and,
during backward, replays only the producers of the values autograd
consumes, so saving a tensor eliminates every producer needed only for
it. A planner selects the saved tensors from measured recomputation
costs under a single limit on step-peak memory. On Qwen3.8-27B and Qwen3.6-35B-A3B, replay alone reduces step time by 3.0\% and 1.7\% versus full checkpointing at no greater memory cost; memory-aware planning raises these gains to 6.3\% and 13.9\% under a 90\%-of-device-memory limit. The evaluated hand-tuned policies fall short of this memory--recomputation frontier and, on Qwen3.6-35B-A3B, run out of memory at larger batches. Over 500 steps on real data, \sys reduces training time by 8--11\%; on Qwen3.5-397B, it is 2.7\% faster than a model-specific hand-tuned policy while using 14\% less memory.

\end{abstract}


\maketitle

\section{Introduction}
\label{sec:intro}

\begin{figure}[t]
\centering
\includegraphics{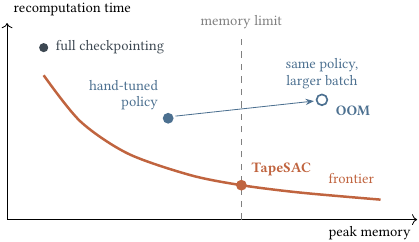}
\caption{Schematic of the peak-memory--recomputation-time trade-off in activation checkpointing. Full checkpointing and a hand-tuned policy lie above the Pareto frontier; \sys targets a frontier point under the memory limit.}
\label{fig:intro-frontier}
\end{figure}

\begin{table*}[t]
\centering
\footnotesize
\caption{Comparison of activation-checkpointing mechanisms: who decides what to save, what is recomputed, how time and memory are estimated, and what each requires.}
\label{tab:mechanisms}
\begin{tabular}{@{}>{\raggedright\arraybackslash}p{0.14\textwidth}>{\raggedright\arraybackslash}p{0.13\textwidth}>{\raggedright\arraybackslash}p{0.25\textwidth}>{\raggedright\arraybackslash}p{0.21\textwidth}>{\raggedright\arraybackslash}p{0.17\textwidth}@{}}
\toprule
\textbf{Mechanism} & \textbf{Who decides what to save} & \textbf{What is recomputed} & \textbf{Time and memory estimates} & \textbf{Requires} \\
\midrule
Full checkpointing & nobody (block inputs only) & the whole block, from its input, in program order & none & eager execution \\
PyTorch SAC & a hand-written policy & the whole block, from its input, except operators whose outputs are saved & none & eager execution \\
\texttt{torch.compile} memory budget & automatic & operators the partitioner does not save, inside the compiled backward & fixed rules, FLOPs or compile-time timing; saved bytes & a captured graph \\
Static and trace-based planners & automatic & the operators of a planned schedule & modeled or profiled time; modeled memory & a known graph or trace, and a runtime that executes the schedule \\
\sys & automatic & only the producers of values the backward pass reads, starting from the nearest saved tensors & measured time; the step's peak memory, from measured must-only steps & eager execution \\
\bottomrule
\end{tabular}
\end{table*}

Training large models relies on several forms of parallelism.
Data-parallel sharding distributes parameters, gradients and optimizer
state across ranks~\cite{zero,fsdp}, while tensor, sequence, context and
expert parallelism partition the computation and, with it, the
activations held by each rank~\cite{shoeybi2019megatron,li2023sequence,liu2024ringattention,lepikhin2021gshard}. Even with all of
them enabled, activation checkpointing~\cite{chen2016} remains necessary:
parallelism reduces per-rank activation memory, but not enough to fit,
so intermediate tensors are discarded after the forward pass and
recomputed when the backward pass requires them. Full
checkpointing reruns each block's forward pass during backward, adding
roughly one forward pass per step; selective activation checkpointing
(SAC)~\cite{megatron-selective} instead saves the intermediates that are
cheap to store and expensive to regenerate, and recomputes the rest.

Each choice of what to save determines a point in a two-dimensional
space: the peak memory of the rank and the recomputation time added to the
backward pass. The objective is the Pareto frontier of this space---for
any amount of available memory, the choice that minimizes recomputation
(Figure~\ref{fig:intro-frontier}). In practice, however, SAC is
configured with a fixed policy, hand-tuned for a particular model,
sequence length and kernel set. Such a policy yields one operating point
per configuration and offers no means of trading memory for
recomputation time; when the model, sequence length or kernels change,
that point moves in unplanned ways and may run out of memory (OOM). When a policy fails to carry over, the fallback is
full checkpointing, which forgoes the trade altogether
(\S\ref{sec:eval-frontier}).

The automatic option built into PyTorch is the memory budget of
\texttt{torch.compile}'s partitioner~\cite{pytorch2}. It handles time and
memory in separate stages rather than optimizing them jointly: ban rules reduce each
operator's recomputation time to zero or infinity, a minimum cut then
selects saved tensors by bytes alone, and when the result exceeds the
budget an additive knapsack decides which banned operators to re-allow.
This pricing is accurate only if every allowed recomputation fuses into
a backward kernel for free, which does not hold for the custom kernels and unfusable
operators that produce many of a large model's activations; the budget,
moreover, bounds saved bytes rather than the step's peak
(\S\ref{sec:bg-compile}). In real training workloads, graph capture also fails before
partitioning, because TorchDynamo encounters graph breaks.

Eager-mode checkpointing avoids graph capture, but still leaves two problems unresolved.
First, the save set must be chosen by hand: PyTorch's selective
checkpointing executes a hand-written policy. Second, PyTorch's checkpointing recomputes each block as a whole, in program order. When backward first requests a saved tensor, it reruns the block's forward pass, skipping calls only if their outputs were cached. Consequently, saving a tensor avoids recomputing its own operator but not its upstream producers. Operators whose outputs are never consumed by backward also run again, and recomputed values remain live together instead of being materialized only when needed.

Avoiding this unnecessary work requires two pieces of information: which forward values the backward pass consumes, and which operators produce those values. We found that eager PyTorch already exposes both through the extension points used by its selective checkpointing mechanism.

The
dispatcher observes every operator a block executes, and autograd's
saved-tensor hooks report which forward values the backward pass
may consume. Accordingly, \sys uses them to record each checkpointed block's forward
pass, as it executes, into a tape in static single-assignment (SSA) form.
During backward it replays only the producers of the values autograd
requests, when it requests them, stops at saved tensors, and frees each
replayed value after its last use. Recomputation thus becomes
tensor-level: what is replayed is determined by the tensors autograd
consumes, not by the operators of the block, and saving a tensor
eliminates the recomputation of every producer needed only for it. Moreover, no changes to the model's forward code are needed; kernels that bypass the dispatcher require only one-time registration as custom operators (\S\ref{sec:visibility}).

\begin{figure*}[t]
\centering
\subcaptionbox{Qwen3.8-27B}{\includegraphics{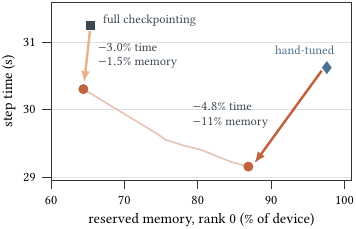}}\hfill
\subcaptionbox{Qwen3.6-35B-A3B}{\includegraphics{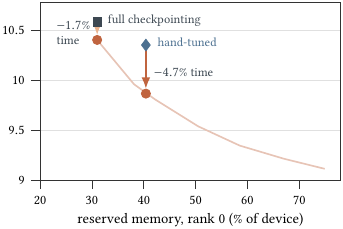}}\hfill
\subcaptionbox{Qwen3.5-397B}{\includegraphics{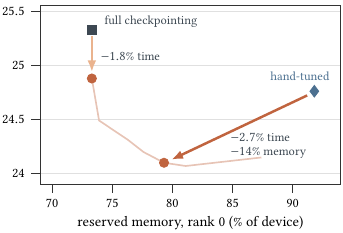}}
\caption{Measured step time and reserved memory under PyTorch SAC and \sys. Light arrows keep the save set of full checkpointing and replace block recomputation with replay; dark arrows replace the hand-tuned policy with a planned save set at no more memory. Faint lines are the \sys frontiers.}
\label{fig:intro-measured}
\end{figure*}

On top of replay, a planner chooses what to save, and two properties
make this choice difficult. The value of saving a tensor depends
on what else is saved, because recomputation chains overlap, so
candidates must be evaluated together on the graph rather than summed one
by one; and the step's peak is set in the backward pass, largely by
model states, communication buffers and allocator fragmentation rather
than by activations. \sys therefore prices
candidates on the recorded graph, with recomputation costs measured on
the actual kernels, and constrains the plan by the step's peak as the
caching allocator sees it. The user specifies a single memory limit, as a
fraction of device memory; sweeping it traces an empirical frontier.

We evaluate \sys on Qwen3.8-27B and Qwen3.6-35B-A3B and validate it on
Qwen3.5-397B. Two findings stand out (Figure~\ref{fig:intro-measured}). First, tensor-level replay
recomputes less than block recomputation: it replays only the minimal subgraph
that the consumed values need, skipping dead nodes (operators whose
outputs no consumed value depends on) and stopping at saved tensors.
With the save set of full checkpointing, where little is saved, it shortens the
step by 1.7--3.0\% at the same memory. Replaying values on demand can
also lower peak memory: reserved memory is 1.5\% lower on the 27B
model with the save set of full checkpointing, and 2.1\% lower on
Qwen3.5-397B with that of its hand-tuned policy.
Second,
automatic planning replaces the hand-tuned policy: because saving a
tensor cuts off the replay of its producers, the planner prices each
candidate by the subgraph it removes, and as the memory limit varies, its
plans trace an empirical frontier of diminishing returns; at 90\% of device memory
they shorten the step by 6.3\% and 13.9\% relative to full checkpointing. The hand-tuned
policies we evaluate lie off this frontier, by up to 5.1\% at the same
memory, and
every hand-tuned policy we evaluate on the 35B model runs out of memory
when the batch grows, while \sys at the same memory limit keeps its gain. Over 500 steps on real data, \sys
shortens training by 8--11\%, and on Qwen3.5-397B it is 2.7\% faster than its hand-tuned
policy while using 14\% less memory.

Table~\ref{tab:mechanisms} contrasts \sys with existing mechanisms, including static and trace-based planners (\S\ref{sec:bg-related}).

Overall, we summarize our main contributions as follows:
\begin{itemize}
\item \textbf{Automatic solving of the save set.} A hand-tuned policy fixes
which tensors to keep, so its memory moves with the configuration and can
run out. \sys needs only a memory limit, as a fraction of device memory:
its planner solves for the saved tensors automatically and solves again
when the configuration changes, and varying the limit traces the
memory--recomputation frontier (\S\ref{sec:planning},
\S\ref{sec:eval-frontier}).
\item \textbf{Recompute only what the backward pass needs, when it needs it.}
PyTorch's checkpointing recomputes the whole block from its input as soon
as backward requests its first saved tensor. \sys recomputes at the
tensor level instead, which brings two gains. First, a value is recomputed only when
autograd requests it, starting from the nearest saved tensors rather than
from the block's input, and is freed after its last use; recomputed
values no longer stay alive together, which can lower the step's peak
memory. Second, \sys skips dead nodes: saving a tensor makes every
producer needed only for it unnecessary, and some operators, such as the
down projection of a multilayer perceptron (MLP), are recomputed by
PyTorch only to re-register their saved inputs. Skipping them shortens the step
at no memory cost, even with the save set of full checkpointing. \sys
reproduces the recomputed values bitwise for any plan (\S\ref{sec:tape},
\S\ref{sec:eval-exact}, \S\ref{sec:eval-replay}).
\item \textbf{Planning on measured costs and the step's peak.} Under replay,
saving a tensor eliminates every producer needed only for it, so the
values of candidates do not add up; and the step's peak is set largely by
model states, communication buffers and fragmentation rather than by
activations. \sys therefore evaluates candidates jointly on the recorded
graph, prices them with measured kernel times, and constrains the step's
peak as the allocator sees it rather than the saved bytes
(\S\ref{sec:planning}, \S\ref{sec:eval-frontier}).
\item \textbf{Directly usable in eager PyTorch.} The dispatcher and the
saved-tensor hooks of eager PyTorch already expose the executed operator
graph and the values the backward pass consumes. Building on them, \sys
needs no graph capture, no change to PyTorch and no edit to the model's forward
code; a kernel that bypasses the dispatcher becomes replayable
after a one-time registration (\S\ref{sec:overview}, \S\ref{sec:visibility}).
\end{itemize}

\section{Background and Related Work}
\label{sec:sota}

This section first formalizes the checkpointing problem
(\S\ref{sec:bg-frontier}). It then examines selective checkpointing in
PyTorch (\S\ref{sec:bg-eager}) and the memory budget of
\texttt{torch.\allowbreak compile} (\S\ref{sec:bg-compile}), stating their
limitations one at a time. Finally, it reviews other rematerialization
systems (\S\ref{sec:bg-related}) and identifies the remaining gaps
(\S\ref{sec:bg-gap}).

\subsection{SAC as a Pareto Problem}
\label{sec:bg-frontier}

Activation checkpointing is applied to a module, typically a transformer
layer; we call each forward call of such a module a \emph{checkpointed
block}. For a block, let $G$ be its forward dataflow graph, $L$ its
inputs, including its parameters, and $P$ its \emph{demand set}, the
forward values that the backward pass may consume. A \emph{save set} $S$ is the set of values saved across the
forward/backward boundary; every other value in $P$ is regenerated
during backward from $S \cup L$. A choice of save sets for all blocks
determines two quantities: $T$, the recomputation time added to the
backward pass, and $M$, the peak device memory of the training step.
The goal is the Pareto frontier of these choices: for a given memory
limit, the feasible save sets with minimal $T$. Neither quantity is a
sum over individual values, as the rest of this subsection explains.

Which calls must be recomputed depends on $S$. To regenerate a value in
$P$ that is not saved, its producer must run, and so must the
producers of that call's inputs, and so on back, stopping at inputs that
are saved or are block inputs, that is, values in $S \cup L$. Every
other call need not be recomputed; we call it a \emph{dead node}. A call
is dead in two cases. In the first, nothing the backward pass reads
depends on its output: for example, a matrix multiplication whose output
only enters an addition, since the gradients of the multiplication need
its inputs and those of the addition need neither operand. In the
second, its output serves only to compute saved values: if
$a = \phi(x)$ and $b = \psi(a)$, and the backward pass reads only $b$,
saving $b$ makes both $\phi$ and $\psi$ dead.

A save set therefore removes the recomputation of every call that it
makes dead, and $T$ is the time of the calls that remain. Whether a call
is dead depends on the set as a whole. The time that a save set removes
is therefore not the sum of the times that its values would remove one
at a time, and once some values are saved, saving one more removes
a different amount of time than it would have at the start. We define the
\emph{return} of saving a value as the recomputation time it removes,
given the values already saved, divided by the bytes it keeps. In the
example above, saving $a$ first removes $f$; but once $b$ is
saved, $f$ is already dead, and saving $a$ removes nothing.
Conversely, if $a$ also feeds a value $c = k(a)$ that the backward pass
reads, saving $b$ first removes only $g$, because $c$ still needs
$f$; once $c$ is saved, saving $b$ removes both $g$ and $f$. The
return of a value thus changes as the save set grows, and the save set
must be evaluated as a whole.

Memory does not add up either. At the moment of the peak $M$, the device
holds much more than saved activations: model states (parameters,
gradients and optimizer state), communication buffers, allocator
fragmentation, and the temporary tensors of the block being recomputed.
A saved value raises $M$ only if it is still alive at that moment; a
value already freed by then does not. Whether saving a value raises
the peak, and by how much, therefore cannot be read from its size alone;
it depends on the step as it actually runs.

\subsection{Selective Checkpointing in PyTorch}
\label{sec:bg-eager}

Selective checkpointing in PyTorch works in two steps. In the forward
pass, a dispatch mode (\texttt{TorchDispatchMode}) intercepts every
operator call in the block and asks a hand-written policy
(\texttt{create\_selective\_checkpoint\_contexts}) whether to cache the
call's output. In the backward pass, when autograd first needs one of
the block's saved values, the block's forward pass is run again from
its input: cached calls return the outputs saved in the forward pass,
and all other calls execute again. We call this way of recomputing
\emph{block recomputation}. The design has three limitations.

\textbf{Limitation 1: The save set is not solved automatically.} The
policy must be written by hand, deciding operator by operator which
outputs to cache. The annotation library
\texttt{torch\_remat}~\cite{torch-remat} replaces the per-operator
policy with save or recompute annotations on regions of the model's
source code, but the choice is still made by hand, and the regions
marked for recomputation are still rerun in full. A hand-written policy
fixes which tensors are kept, not how much memory they take: when the
configuration changes, its memory changes with it and can run out
(\S\ref{sec:eval-frontier}), and every new configuration or memory
limit requires rewriting the policy.

\textbf{Limitation 2: Recomputation always starts from the block's
input.} Block recomputation runs from the first operator onward and
regenerates, in one pass, every value that is not cached; all of them
then stay in memory until autograd reads them. Yet the saved values
already provide starting points inside the block: each value could be
regenerated from the nearest saved values upstream of it, only when
autograd needs it, and freed after its last use. Values that autograd
reads only near the end of the block's backward pass would then not
occupy memory from its start, which can lower the peak
(\S\ref{sec:bg-frontier}).

\textbf{Limitation 3: Dead nodes are recomputed.} Block recomputation runs
every call that is not cached, whether or not the backward pass needs
its result, so it recomputes both kinds of dead node defined in
\S\ref{sec:bg-frontier}. Calls that the backward pass does not need at
all still run: on Qwen3.6-35B-A3B, for example, block recomputation reruns
the experts' down projection in every mixture-of-experts (MoE) layer,
although the backward pass does not read its output
(\S\ref{sec:eval-replay}). Calls needed only by cached values also
still run, because block recomputation starts from the block's input
(Limitation 2); caching a value therefore removes only the call that
produces it, not the calls upstream of it. The \texttt{early\_stop}
option removes only the calls after the last value that the backward
pass needs has been regenerated, at the end of the block, and no dead node before that
point.

Limitations 2 and 3 share a cause: block recomputation has neither the block's
dataflow graph $G$ nor the set $P$ of values the backward pass consumes
(\S\ref{sec:bg-frontier}), so it can tell neither where a value could
be regenerated from nor whether a call is needed at all. \emph{Replay},
which executes on demand only the producers of the consumed values and
stops at saved values, needs both; \S\ref{sec:tape-exec} formalizes
the difference in cost.

\subsection{The Memory Budget of \texttt{torch.compile}}
\label{sec:bg-compile}

The memory budget of \texttt{torch.compile} chooses the save set
automatically. Its partitioner decides what to save on the joint
forward--backward graph~\cite{pytorch2}. It was designed for a compiler
that fuses cheap operators into backward kernels, and it assumes that
recomputing such operators is free and that other operators should not be
recomputed at all. By default, it therefore does not recompute matrix
multiplications, attention or custom operators, and among the remaining
operators it uses a minimum cut, as in fusion-aware
checkpointing~\cite{fusionaware}, to pick the save set with the fewest
saved bytes, weighted by heuristics.

The memory budget (\texttt{activation\_memory\_budget}, between 0 and 1)
asks for fewer saved bytes than this default. The partitioner first
relaxes some of its restrictions; if the saved bytes are still above the
budget, a 0-1 knapsack that weighs each operator's size against its
estimated runtime chooses which expensive operators to recompute, and the
bytes-only cut is rerun with those operators
allowed~\cite{pytorch-partitioner}. The budget mode thus applies a cut
built for free recomputation to exactly the case where recomputation is
not free: time enters only through the knapsack, one operator at a time.
The design has four limitations.

\textbf{Limitation 4: The graph of real training is not captured
whole.} The budget applies only to a graph that \texttt{torch.compile}
has captured. On the three models we train, the graph breaks before the
partitioner runs, so the budget cannot be applied at all
(\S\ref{sec:eval-compile}). Where a break splits a model into several
graphs, each graph is partitioned on its own, and no decision covers the
training step as a whole.

\textbf{Limitation 5: Recomputation time is not measured on the training
step.} By default, the knapsack uses an operator's count of
floating-point operations (FLOPs) in place of its time. FLOPs cannot
stand in for time: how long a kernel takes in training depends on much
more than its arithmetic, for example on the machine, on some devices
being slower than others, and on communication that overlaps the kernel
and slows it down. The other estimator does time operators, but during
compilation, outside the training step. Recomputation time has to be
measured on the training step itself.

\textbf{Limitation 6: The memory limit is not the step's peak.} The
budget is a fraction of the bytes that one graph saves for backward by
default, not an amount of device memory. Whether a step fits, however,
depends on the peak memory of the whole step (\S\ref{sec:bg-frontier}),
and much of that peak lies outside the budget: model states,
communication buffers, the allocator's fragmentation and the activations
of other graphs. The peak also depends on when values are alive, not
only on how many bytes are saved. It falls inside the backward window of
one block: by then, the blocks whose backward pass has finished have
freed their saved values, and the block itself adds the temporaries of
its own recomputation. A count of saved bytes sees neither. A user
therefore cannot derive the budget from the memory available; on
controlled models, even a budget of zero leaves the peak unchanged
(\S\ref{sec:eval-compile}). Planning needs the peak measured on the
training step, including where in the backward pass it falls.

\begin{figure}[t]
\centering
\begin{tikzpicture}[font=\scriptsize,
    op/.style={draw, rounded corners=2pt, minimum height=4.2mm, inner sep=2pt},
    re/.style={op, fill=tsFull!8},
    sv/.style={op, draw=tsOurs, fill=tsOurs!15, thick},
    dd/.style={op, dashed, draw=gray!70, text=gray},
    arr/.style={-{Latex[length=1.2mm]}},
    darr/.style={arr, dashed, gray!70},
    lab/.style={anchor=west, align=left}]
  \node[re, minimum width=12mm] (ma) at (0,0) {matmul};
  \node[re, minimum width=15mm] (qa) at (2.3,0.25) {query copy};
  \node[sv, minimum width=15mm] (ga) at (2.3,-0.25) {gate};
  \draw[arr] (ma.east) -- (qa.west);
  \draw[arr] (ma.east) -- (ga.west);
  \node[lab] at (3.3,0) {(a) Save the gate:\\matmul recomputed, 2.0\,ms removed.};
  \node[re, minimum width=12mm] (mb) at (0,-1.1) {matmul};
  \node[sv, minimum width=15mm] (qb) at (2.3,-0.85) {query copy};
  \node[re, minimum width=15mm] (gb) at (2.3,-1.35) {gate};
  \draw[arr] (mb.east) -- (qb.west);
  \draw[arr] (mb.east) -- (gb.west);
  \node[lab] at (3.3,-1.1) {(b) Save the query copy:\\matmul recomputed, 1.1\,ms removed.};
  \node[dd, minimum width=12mm] (mc) at (0,-2.2) {matmul};
  \node[sv, minimum width=15mm] (qc) at (2.3,-1.95) {query copy};
  \node[sv, minimum width=15mm] (gc) at (2.3,-2.45) {gate};
  \draw[darr] (mc.east) -- (qc.west);
  \draw[darr] (mc.east) -- (gc.west);
  \node[lab] at (3.3,-2.2) {(c) Save both:\\matmul dead, 6.0\,ms removed.};
  \node[sv, minimum width=5mm] (k1) at (0,-3.1) {};
  \node[lab] at (k1.east) {saved for backward};
  \node[re, minimum width=5mm] (k2) at (2.55,-3.1) {};
  \node[lab] at (k2.east) {recomputed};
  \node[dd, minimum width=5mm] (k3) at (4.75,-3.1) {};
  \node[lab] at (k3.east) {dead, not recomputed};
\end{tikzpicture}
\caption{A joint effect in a full-attention block of Qwen3.6-35B-A3B.
One matrix multiplication produces both the query and the output gate;
it is dead only when both are saved. Times are the replay time removed per
block.}
\label{fig:joint}
\end{figure}
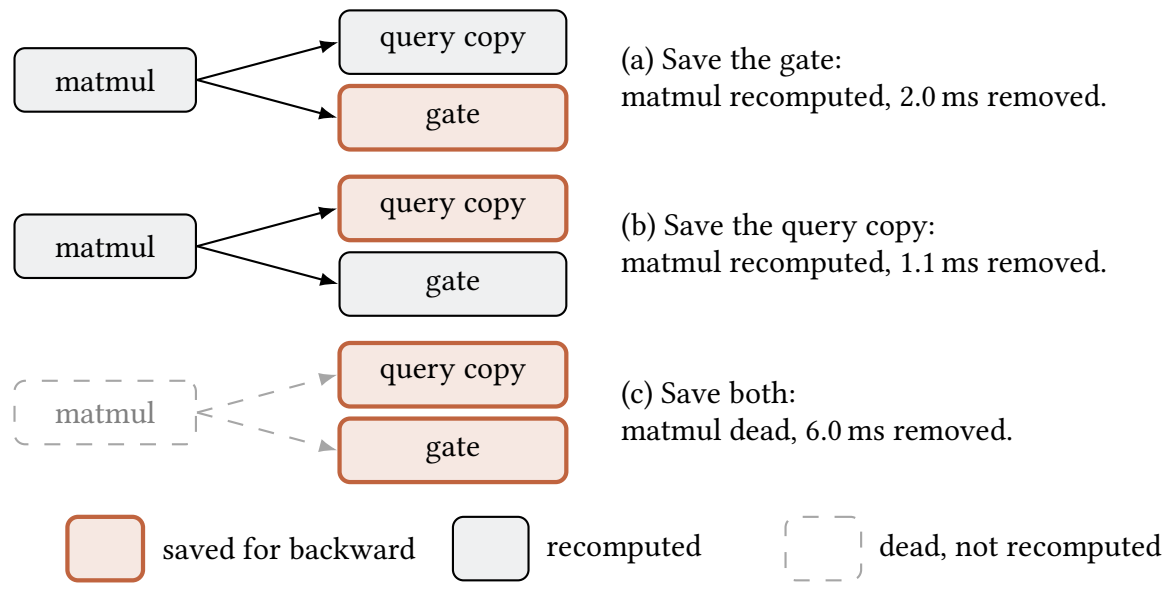

\textbf{Limitation 7: Dead nodes and joint effects are not captured.}
The knapsack scores each operator by its own estimated time and adds the
scores up. The return of saving a value, however, is neither limited
to the call that produces it nor fixed (\S\ref{sec:bg-frontier}): it
includes the time of the upstream calls that saving the value makes
dead, and it depends on what else is saved. Two values saved
together can remove more time than the sum of what each removes alone.
In Figure~\ref{fig:joint}, from Qwen3.6-35B-A3B, saving the query copy
or the gate alone removes 1.1 or 2.0\,ms of replay time per block, whereas
saving both removes 6.0\,ms, because the matrix multiplication that
produces them becomes dead. A fixed, additive score can express
neither effect, and the cut, although it sees the whole graph, counts
only bytes. No part of the partitioner therefore evaluates the time
that the finally chosen save set removes.

Limitations 5--7 matter most when memory is scarce enough that matrix
multiplications and custom operators must be recomputed, which is the
regime of the models studied here; \S\ref{sec:eval-compile} quantifies
their effect on Qwen3-8B, which the compiler can capture, and on
controlled models.

\subsection{Other Rematerialization Systems}
\label{sec:bg-related}

\emph{Static-graph planners.} Revolve and sublinear-memory checkpointing
place checkpoints on chains~\cite{revolve,chen2016}, and Rotor solves
heterogeneous chains by dynamic programming~\cite{rotor}. Checkmate,
Moccasin and XEngine schedule rematerialization on general graphs with
integer or constraint
programming~\cite{checkmate,moccasin,schuler2022xengine}, while MODeL
reorders operators and assigns tensor addresses to lower the
peak~\cite{model}. For large language models, Yuan et al.\ enumerate
save sets within a transformer layer to obtain a memory--computation
frontier and search it jointly with the parallel
configuration~\cite{yuan2024activation}. These planners assume a known
graph or layer structure and a runtime that executes the chosen
schedule. In contrast, \sys records the concrete graph of each eager
invocation and keeps autograd's backward.

\emph{Trace-based and runtime planners.} Rockmate traces the forward pass
on an example input, partitions the trace, and generates a module that
drives the backward pass according to its schedule~\cite{rockmate};
HiRemate extends this approach with hierarchical planning~\cite{hiremate}.
Mimose predicts per-layer memory online and caches a checkpoint plan for
each input size~\cite{liao2023mimose}; however, it plans whole layers
and recomputes each layer from its input (Limitations 2 and 3). \sys, by
contrast, needs no ahead-of-time trace, keeps autograd's backward, and
records and replays at the tensor level.

\emph{Estimation-based planners.} Mist traces a symbolic-shape graph,
constructs a backward graph to estimate time and memory, and co-optimizes
activation checkpointing with the parallel configuration~\cite{mist}.
PyTorch's experimental SAC tools estimate operator costs on fake tensors,
either with a roofline model of peak device throughput or by benchmarking
each operator that the dispatcher sees~\cite{pytorch-sac-ilp}. Under an
estimated peak, they decide which modules to checkpoint and how much of
their activations to discard; within each module, a knapsack whose value
is the sum of operator runtimes selects the operators to save. PyTorch's
SAC then executes the result with block recomputation and thus inherits
Limitations 2 and 3. Like the memory budget, these planners estimate
costs rather than measure them on the training step (Limitation 5),
plan under an estimated peak rather than the measured one
(Limitation 6), and score each operator by its own runtime
(Limitation 7).
In contrast, \sys measures the concrete invocation, evaluates saved
tensors jointly on the recorded graph, and replays rather than
recomputing whole blocks.

\emph{Reactive rematerialization.} DTR and Capuchin evict tensors under
memory pressure and regenerate them from recorded lineage
\cite{dtr,capuchin}, and Coop evicts contiguous tensors together to limit
the allocator fragmentation that tensor-wise eviction causes~\cite{coop}.
These systems decide at run time what to evict. \sys instead plans a
repeatable operating point from measured costs under the step's peak.

\subsection{Summary}
\label{sec:bg-gap}

Table~\ref{tab:mechanisms} contrasts these mechanisms. PyTorch's SAC
leaves the choice to a hand-written policy (Limitation 1) and recomputes
whole blocks (Limitations 2 and 3). The compiler chooses
automatically, but only on captured graphs (Limitation 4), without
measuring time or the peak on the training step (Limitations 5 and 6),
and without knowing the return of what it saves (Limitation 7).
Rematerialization planners require a static graph or trace, estimate
costs rather than measure them, take over the backward pass, or plan
whole layers. To the best of our knowledge, no prior mechanism for eager
PyTorch chooses save sets at the tensor level from measured kernel costs
under the step's peak memory while replaying only the producers that
autograd's demand requires.

\section{Methodology}
\label{sec:method}

\sys has two layers. The recording layer captures each checkpointed
block as a tape of the operator calls that actually ran and, during
backward, replays only what autograd needs; whichever tensors are
saved, backward receives the values the forward pass produced. The
planning layer chooses those tensors: under a memory limit, it finds the
plan with the least recomputation time, from costs and peak memory
measured on the training step itself. Recording makes every choice safe;
planning makes the choice good. Sections~\ref{sec:tape}
and~\ref{sec:visibility} describe recording, and
Section~\ref{sec:planning} planning.

\subsection{Overview}
\label{sec:overview}

Eager PyTorch routes every ATen operator, every functional collective and
every library-registered custom operator through the dispatcher, where a
\texttt{TorchDispatchMode} observes each call with its concrete arguments
and results; PyTorch's own SAC is built on this mechanism. In the same
forward pass, \texttt{saved\_tensors\_hooks} report which tensors autograd
saves for backward. Together they provide, for each block invocation, the
forward graph $G$ and the demand set $P$ of \S\ref{sec:bg-frontier}, for
the computation that actually executed and without ahead-of-time capture,
except for fused kernels that are not registered custom operators
(\S\ref{sec:visibility}).

\begin{figure*}[t]
\centering
\begin{tikzpicture}[font=\scriptsize,
    op/.style={draw, rounded corners=2pt, minimum height=4.6mm, inner sep=2pt},
    re/.style={op, fill=tsFull!8},
    sv/.style={op, draw=tsOurs, fill=tsOurs!15, thick},
    dd/.style={op, dashed, draw=gray!70, text=gray},
    leaf/.style={op, fill=white},
    arr/.style={-{Latex[length=1.2mm]}},
    darr/.style={arr, dashed, gray!70},
    val/.style={inner sep=1pt},
    cell/.style={draw, rounded corners=2pt, minimum height=8mm, minimum width=16mm, align=center},
    note/.style={align=flush left, text width=16.5mm, anchor=north, inner sep=1pt, yshift=-1mm},
    lab/.style={anchor=west, align=left}]
  \node[anchor=west, font=\footnotesize\bfseries] at (-0.25,1.75) {(a) One block: record in forward, replay in backward};
  \node[leaf] (x) at (0,0) {$x$};
  \node[re]   (norm) at (1.1,0) {norm};
  \node[dd]   (mg) at (2.75,0.55) {mm \#0 (gate)};
  \node[re]   (mu) at (2.75,-0.55) {mm \#1 (up)};
  \node[sv, minimum width=5mm] (g) at (4.2,0.55) {$g$};
  \node[re]   (silu) at (5.15,0.55) {silu};
  \node[re]   (mul) at (6.35,0) {mul};
  \node[dd]   (md) at (7.85,0) {mm \#2 (down)};
  \node[gray, align=left, anchor=west] (y) at (9.0,0) {$y$: not saved};
  \draw[arr] (x) -- (norm);
  \draw[darr] (norm.east) -- node[val, above left, pos=0.45, text=black] {$h$} (mg.west);
  \draw[arr] (norm.east) -- (mu.west);
  \draw[darr] (mg) -- (g);
  \draw[arr] (g) -- (silu);
  \draw[arr] (silu.east) -- node[val, above right, pos=0.3] {$s$} (mul.west);
  \draw[arr] (mu.east) -- node[val, below, pos=0.5] {$u$} (mul.west);
  \draw[arr] (mul) -- node[val, above] {$m$} (md);
  \draw[darr] (md) -- (y.west);
  \draw[-{Latex[length=1.5mm]}, thick, tsOurs!80] (8.6,-1.1) -- (0.1,-1.1);
  \node[align=center, anchor=north] at (4.35,-1.15)
    {backward unpacks $h, u, s, m$ $\Rightarrow$ recompute norm, mm \#1, silu, mul from $x$ and $g$;\\
     mm \#0 and mm \#2 are not recomputed};
  \node[anchor=west, font=\footnotesize\bfseries] at (10.3,1.75) {(b) Planning over training steps};
  \node[cell] (s1) at (11.15,0.3) {warm-up\\ $W$ steps};
  \node[cell, fill=tsFull!8] (s2) at (12.95,0.3) {profile\\ $R$ steps};
  \node[cell, fill=tsOurs!15, draw=tsOurs, thick] (s3) at (14.75,0.3) {solve\\ one step};
  \node[cell] (s4) at (16.55,0.3) {train\\ later steps};
  \draw[arr] (s1) -- (s2); \draw[arr] (s2) -- (s3); \draw[arr] (s3) -- (s4);
  \node[note] at (s1.south) {warm-up with mandatory values only};
  \node[note] at (s2.south) {time the recomputed nodes of one block per type; measure the must-only step peak};
  \node[note] at (s3.south) {rank 0 solves under the memory limit; broadcast};
  \node[note] at (s4.south) {reuse the plan of the token bucket; solve again for a new bucket};
  \draw[arr, tsOurs!80, thick, rounded corners=3pt] (s3.north) -- (14.75,1.25) -- (4.2,1.25)
    node[pos=0.62, above, text=black] {plan: save output 0 of mm \#0} -- (g.north);
  \node[sv, minimum width=5mm] (k1) at (0,-2.3) {};
  \node[lab] at (k1.east) {saved for backward};
  \node[re, minimum width=5mm] (k2) at (2.6,-2.3) {};
  \node[lab] at (k2.east) {recomputed};
  \node[dd, minimum width=5mm] (k3) at (4.5,-2.3) {};
  \node[lab] at (k3.east) {dead, not recomputed};
  \node[leaf, minimum width=5mm] (k4) at (7.2,-2.3) {};
  \node[lab] at (k4.east) {block input};
\end{tikzpicture}
\caption{Execution and planning in \sys: (a) recording and replay of a
gated MLP block under a plan that saves $g$; (b) profiling and
solving over training steps.}
\label{fig:overview}
\end{figure*}
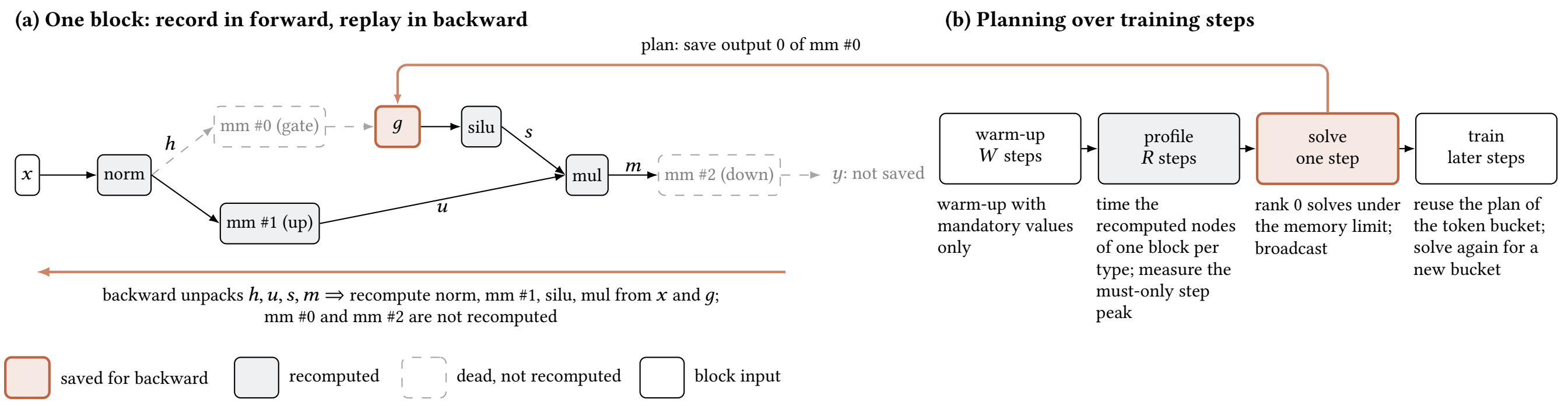

Figure~\ref{fig:overview} illustrates recording and planning on a gated MLP block,
and Algorithm~\ref{alg:step} gives one training step. To adopt \sys, the
training code swaps its checkpoint wrapper for
\sys's, and a model adapter reports each block's region
type, the kind of block that the planner budgets separately
(\S\ref{sec:planning-model}), and its token count. In every step, each block
records its tape and saves the planned and mandatory values during
forward, and the executor replays on demand during backward
(\S\ref{sec:tape}). Plans come from profiling a few early steps
(Figure~\ref{fig:overview}b): after $W$ warm-up steps, the next $R$ steps
($W = R = 2$ in all experiments) time the operators
(reference steps) and measure peak memory (sample steps), and rank~0
then solves a plan
for each range of token counts that occurs. Profiling is repeated when
token counts move far from the profiled ones (\S\ref{sec:planning}).

\subsection{Tensor-Level Record and Replay}
\label{sec:tape}

Each checkpointed block runs its forward pass as usual, under two hooks
that eager PyTorch provides. A dispatch mode records every operator call
of the block into a tape. Autograd's saved-tensor hooks see every tensor
that autograd saves for backward: \sys keeps the tensors that must be
saved or that the plan saves, and replaces every other one with a value
identifier, so that, as under full checkpointing, the tensor is freed
after forward. In backward, when autograd unpacks an identifier, \sys
recomputes that value from the tape, running only the operators it
depends on. The rest of this section describes the tape
(\S\ref{sec:tape-record}), the values that forward keeps
(\S\ref{sec:tape-pack}), and how replay recomputes the others
(\S\ref{sec:tape-exec}).

\subsubsection{Recording the tape}
\label{sec:tape-record}

While a block runs forward, the dispatch mode records every operator
call, in order, as a node
$(\mathit{op}, \mathit{args}, \mathit{in\_vids}, \mathit{out\_vids})$:
the operator, its arguments with each tensor replaced by its value
identifier, and the identifiers of its tensor inputs and outputs,
written $\mathrm{in}(n)$ and $\mathrm{out}(n)$ below. Every value is
either the output of exactly one node, its producer $\mathrm{prod}(v)$,
or a \emph{leaf}: an input of the block or a parameter, which the block
uses but does not compute. The tape is thus the dataflow graph $G$ of
\S\ref{sec:bg-frontier}, recorded from the calls that actually ran. We
write $V$ for all values and $L \subseteq V$ for the leaves. The block
of Figure~\ref{fig:overview}a gives six nodes; $x$ and the weights are
leaves, and $g$ is produced by mm~\#0.

The tape records how each value was computed, that is, which operator
produced it from which inputs, but not the value's data. Apart from
saved values (\S\ref{sec:tape-pack}), it holds no
reference to any intermediate tensor; such a reference would keep the
tensor alive past forward and defeat checkpointing. Three details keep
this bookkeeping exact. An in-place update creates a new value, so a
node that runs after it sees the updated tensor as a different value
from the original. A view shares memory with its base, so their bytes
are counted once. Parameters are recorded by name and fetched again in
backward, so that under fully sharded data parallelism
(FSDP2~\cite{fsdp}) backward does not keep the unsharded copies of
forward.

\subsubsection{Values that forward keeps}
\label{sec:tape-pack}

Backward does not need every forward value, only those that autograd
saves for its gradient formulas. They form the demand set
$P \subseteq V$ (\S\ref{sec:bg-frontier}). In Figure~\ref{fig:overview}a, $P$ contains $h$, $g$,
$s$, $u$ and $m$: the gradient of silu needs $g$, that of mul needs $s$
and $u$, and so on. It does not contain $y$, which no gradient formula
needs. Autograd hands every value of $P$ to the saved-tensor hooks. For a
value in the save set $S$, the tape holds the tensor; for a leaf, it
holds the block input or the parameter's name; for every other value,
autograd keeps only the identifier, and the tensor is freed when forward
no longer uses it.

A few values must always be saved, because replaying them is unsafe. They
form the \emph{mandatory set}, which every save set contains:
\begin{equation}
S_0 = V_{\mathrm{coll}} \cup V_{\mathrm{dyn}} \cup V_{\mathrm{alloc}},
\qquad S \supseteq S_0.
\label{eq:mandatory}
\end{equation}
$V_{\mathrm{coll}}$ holds the outputs of collective communication, which
replay would have to communicate again. $V_{\mathrm{dyn}}$ holds outputs
whose size depends on the data, such as that of \texttt{nonzero}, which
replay could allocate only after the CPU waits for the GPU.
$V_{\mathrm{alloc}}$ holds buffers filled by a kernel that the tape cannot
see (\S\ref{sec:visibility}), which replay would recreate empty.
A block or step that saves only $S_0$ is \emph{must-only}.

\label{sec:tape-granularity}
The rest of $S$ comes from a plan, which the planner of
\S\ref{sec:planning} solves once and later steps apply. Those steps'
tensors do not exist yet when the plan is solved, so the plan names each
value by its position in the block:
\begin{equation}
\mathrm{coord}(v) = (\mathit{op}, k, j),
\label{eq:coord}
\end{equation}
output~$j$ of the $k$-th call of operator $\mathit{op}$ in the block,
counting from~0. Every step runs the same block code, so a coordinate
points to the same place in every step. Recording applies the plan: for
each operator call, it counts how many times that operator has already
been called in the block, which gives $k$, and saves output~$j$ if
$(\mathit{op}, k, j)$ is in the plan. With the plan
$\{(\texttt{mm}, 0, 0)\}$ of Figure~\ref{fig:overview}a, the first
\texttt{mm} saves its output $g$, and the second and third save nothing.
Because a plan lists outputs rather than calls, it can save one output of
a multi-output kernel and drop another, for example the log-sum-exp of
an attention kernel but not its output.

\subsubsection{Replay}
\label{sec:tape-exec}

Replay starts when autograd first unpacks a value of the block. \sys
first counts, for every value that replay may produce, how many times it
will be read, as the input of another replayed node or by autograd. Then,
for each value that autograd unpacks, it returns the value if it is
available, that is, a leaf, a saved value or a value replayed earlier.
Otherwise, it walks back from the value through its producers, collecting
every node it passes until it reaches available values, and runs the
collected nodes in dependency order. We call the nodes collected for targets
$Q$ with available values $B$ the \emph{producer closure} of $Q$, written
$\closure_B(Q)$. In Figure~\ref{fig:overview}a, backward first asks for $m$. The walk goes
from mul to silu, where it stops at the saved $g$, and to mm~\#1 and norm,
where it stops at the input $x$. These four nodes run, and the values they
produce, $h$, $s$ and $u$, are ready when backward asks for them next.
Replay runs whole calls: if backward needs the output of an attention
kernel whose log-sum-exp is saved, the whole kernel runs again.

Each replayed value is freed when its count reaches zero. A value that
backward needs only late is therefore computed late and freed right after
use, instead of occupying memory from the start of the block's backward
(Limitation~2). Replay never writes into a saved value or a leaf; it
writes into a copy.

Over the whole backward, replay runs only the nodes
\begin{equation}
N_{\mathrm{rp}}(S) = \closure_{S \cup L}(P),
\label{eq:n-rp}
\end{equation}
each at most once. All other nodes never run; they are the dead nodes of
\S\ref{sec:bg-frontier}. In Figure~\ref{fig:overview}a, mm~\#0 never
runs because its only output $g$ is saved, and mm~\#2 because backward
needs none of its outputs.

PyTorch's block recomputation (\S\ref{sec:bg-eager}) instead reruns forward from
the first node. It skips the cached calls and, with \texttt{early\_stop},
stops after the last value it still needs:
\begin{equation}
N_{\mathrm{blk}}(S) = \{\, n_i : \mathrm{out}(n_i) \not\subseteq S,\
  i \le i^{*}(S) \,\},
\label{eq:n-re}
\end{equation}
where $n_i$ is the $i$-th node of the tape and $i^{*}(S)$ is the position
of the last node with an output in $P \setminus S$. With $t(n)$ the measured time of node $n$, the recomputation times of
the two are
\begin{equation}
T_{\mathrm{blk}}(S) = \sum_{n \in N_{\mathrm{blk}}(S)} t(n),
\label{eq:t-re}
\end{equation}
\begin{equation}
T_{\mathrm{rp}}(S) = \sum_{n \in N_{\mathrm{rp}}(S)} t(n).
\label{eq:t-rp}
\end{equation}
Block recomputation runs every node that replay runs, so for any save set that
both can express
\begin{equation}
N_{\mathrm{rp}}(S) \subseteq N_{\mathrm{blk}}(S), \qquad
T_{\mathrm{blk}}(S) - T_{\mathrm{rp}}(S) =
  \sum_{n \in N_{\mathrm{blk}}(S) \setminus N_{\mathrm{rp}}(S)} t(n).
\label{eq:gap}
\end{equation}
In words, replay removes exactly the time of the dead nodes that
block recomputation still runs. Because autograd may ask for only part of $P$,
Eq.~\ref{eq:t-rp} is an upper bound on replay time.

\subsection{Registering Kernels as Custom Operators}
\label{sec:visibility}

The tape sees only operators that go through the dispatcher. Many custom
kernels, such as FlashAttention, the chunked scan of Gated DeltaNet (GDN)
linear attention~\cite{gdn}, fused normalizations and rotary position
embeddings (RoPE), and expert GEMMs (general matrix multiplications),
are called from an \texttt{autograd.Function} rather than registered as
custom operators, so the dispatcher does not capture them.

\label{sec:vis-register}
\sys registers each such kernel once as a custom operator
(\texttt{torch.library.custom\_op}), with the original kernel as its
forward and the original backward as its backward; neither the kernel
nor the model code changes. The tape then records the kernel as one
node, replays it, and can save each of its outputs separately, for
example the attention log-sum-exp without the attention output.
\label{sec:vis-mustsave}

\subsection{Peak-Constrained Planning}
\label{sec:planning}

Given a memory limit $f$, a fraction of device memory per rank, the
planner chooses the values to save that minimize the predicted replay
time while keeping the predicted peak memory of the step within the
limit. It profiles one block of each region type and derives candidate
values from it (\S\ref{sec:planning-model}), computes how much memory the
step leaves for saved values (\S\ref{sec:planning-mem}), splits that
memory among region types and blocks (\S\ref{sec:planning-solve}), and
shares one plan across ranks and steps (\S\ref{sec:planning-keys}).

\begin{algorithm}[t]
\caption{Planning and execution of a \sys training step.}
\label{alg:step}
\small
\begin{algorithmic}[1]
\algtext*{EndIf}
\algtext*{EndFor}
\Require limit $f$; per region type $r$: tape $G_r$, node times $t(n)$ and
  block count $m_r$ (reference steps); $\mathrm{base}(\tau)$ (sample
  steps); plan cache $\Pi$
\Statex \textbf{Plan}
\State $\bar\tau \gets \max_{\text{ranks}} \tau$;\ \ $q \gets \mathrm{bucket}(\bar\tau)$
\If{$\Pi[q] = \emptyset$}
  \Comment{solved on rank~0}
  \State $\Delta \gets \min_{\text{ranks}} \bigl(C - \mathrm{base}(\tau_q^{+})\bigr)$
    \Comment{Eqs.~\ref{eq:step-peak}, \ref{eq:capacity}}
  \For{each region type $r$}
    \State $F_r \gets$ frontier of $\bigl(b(S),\, T_{\mathrm{rp}}(S_0) - T_{\mathrm{rp}}(S_0 \cup S)\bigr)$ by beam search
  \EndFor
  \State pick $(b_r, g_r) \in F_r$ maximizing $\sum_r m_r g_r$ s.t.\ $\sum_r m_r b_r \le \Delta$
  \State $\Pi[q] \gets \{(r, \mathrm{coord}(v)) : v \in S_r\}$; broadcast $\Pi[q]$
\EndIf
\Statex \textbf{Forward}, each block of region type $r$
\State record $G$, $L$, $P$; save $S \gets S_0 \cup \{v : (r, \mathrm{coord}(v)) \in \Pi[q]\}$
\Statex \textbf{Backward}, each unpack of $v \in P$
\If{$v \notin S \cup L \cup E$}
  \Comment{$E$: live replayed values}
  \State run $\closure_{S \cup L \cup E}(\{v\})$ in dependency order; add its outputs to $E$
\EndIf
\State return $v$; free each $u \in E$ with no remaining use
\end{algorithmic}
\end{algorithm}

\subsubsection{Region models from profiling}
\label{sec:planning-model}

Every checkpointed block records its tape in every step
(\S\ref{sec:tape-record}), and the model adapter labels the block with
its region type (\S\ref{sec:overview}). Region types depend on the
model; the models of \S\ref{sec:eval} have five: the text decoder's
linear-attention, full-attention and multi-token-prediction (MTP)
blocks, and the vision and audio blocks.
Types differ in the operators
that dominate their time and in the sizes of the values those operators
produce, so the same saved bytes remove different replay time in
different types. The planner therefore models each type separately and
lets the types compete for one capacity (\S\ref{sec:planning-solve}). The
text decoder's types are \emph{primary}: their token count $\tau$ drives
the memory constraint.

The planner profiles a few early steps (Figure~\ref{fig:overview}b). The
first $W$ steps run must-only. On the next step, each region type records
the tape of one reference block, the first block with the type's most
common operator sequence, together with the size of every value, and
times every node that the block's backward replays with CUDA events. A
second step repeats the timing, and the planner keeps each node's shorter
time, which filters out contention from communication. The first of these
steps also runs must-only to sample memory: it records the peak allocated
memory of the whole step, the device memory outside PyTorch's caching
allocator, and the allocator's fragmentation. Sizes and times are scaled
linearly with the region's token count. Blocks whose operator sequence
differs from the reference save only the mandatory set. A region type is
profiled again when its token count exceeds twice that of its reference
block, and memory is sampled again when the token count moves more than
15\% away from the sampled one.

\label{sec:planning-cands}
We observe that saving values together can remove more replay time than
saving them one at a time. In the full-attention block of
Figure~\ref{fig:joint}, saving the query copy or the gate alone removes
1.1 or 2.0\,ms per block, because the matrix multiplication that produces
both still runs for the other; saving both removes 6.0\,ms, because that
matrix multiplication becomes dead as well. A candidate is therefore one
or more output slots that are saved together. Its cost is the memory it
keeps alive, counted once per storage; its benefit is the reduction in
replay time (Eq.~\ref{eq:t-rp}) that it brings.

\subsubsection{A constraint on the step's peak memory}
\label{sec:planning-mem}

Let $\mathrm{base}(\tau)$ be the peak allocated memory of a must-only
step at primary token count $\tau$, interpolated between memory samples.
For region type $r$, let $b_r$ be the bytes a plan saves per block beyond
$S_0$ and $m_r$ the number of blocks of that type that share the
reference's operator sequence. The plan must satisfy
\begin{equation}
\mathrm{base}(\tau) + \sum_r m_r b_r \le C,
\label{eq:step-peak}
\end{equation}
with capacity
\begin{equation}
C = f \cdot M_{\mathrm{dev}} - \mathrm{non\_torch}
    - \mathrm{margin} - \mathrm{frag},
\label{eq:capacity}
\end{equation}
where $M_{\mathrm{dev}}$ is the total device memory,
$\mathrm{non\_torch}$ is the device memory outside PyTorch's caching
allocator, and $\mathrm{frag}$ is the allocator's fragmentation; the
latter two are measured. The safety reserve $\mathrm{margin}$ is about
2\% of device memory in our experiments. If $\mathrm{base}(\tau)$ alone
exceeds $C$, the plan saves nothing beyond the mandatory set; $f = 0$
forces this must-only plan.

Eq.~\ref{eq:step-peak} constrains the whole step, not saved bytes alone
(Limitation~6): $\mathrm{base}(\tau)$ already contains model states,
communication buffers and the temporaries of must-only replay. It charges
every saved byte at once. Every saved value is still alive when forward
ends, so this is nearly exact early in the backward pass, where the peak
falls on our models (\S\ref{sec:eval-memmodel}), and conservative later,
as blocks free their values. When the reserved memory observed in
training approaches the limit, the planner raises its fragmentation
estimate and solves again, which can only shrink the plan.

When the peak falls later in the backward pass, as it can when
activations are offloaded to host memory, Eq.~\ref{eq:step-peak}
overcharges: blocks whose backward pass has finished have freed their
saved values, and saving a value in the block that sets the peak lowers
that block's replay temporaries. For this case, \sys also provides a
peak-aware mode, whose window model predicts the peak of every backward
window from the memory measured between unpacks on a sample step; the
planner then saves extra values in the block that sets the peak. On our models, the
peak falls at the start of the backward pass, where
Eq.~\ref{eq:step-peak} exceeds it by at most one block's saved bytes
(\S\ref{sec:eval-memmodel}), so our experiments use
Eq.~\ref{eq:step-peak}.

\subsubsection{Solver}
\label{sec:planning-solve}

For each region type, the planner builds an approximate Pareto frontier
of sets $S$ of candidates saved on top of the mandatory set $S_0$, in
bytes per block $b(S)$ and replay time removed per
block $T_{\mathrm{rp}}(S_0) - T_{\mathrm{rp}}(S_0 \cup S)$, by a
beam search. Starting from the empty set, each round extends every set
on the frontier by one candidate, rescores the extension by walking the
tape, and keeps the sets that no other set beats in both bytes and time
removed; at most 24 sets, spread over bytes, go on to the next round, and
a set holds at most 12 candidates. A multiple-choice knapsack then
selects one frontier point $(b_r, g_r)$ per region type, where
$g_r = T_{\mathrm{rp}}(S_0) - T_{\mathrm{rp}}(S_0 \cup S_r)$ is the
replay time its save set $S_r$ removes per block, maximizing
$\sum_r m_r g_r$ subject to $\sum_r m_r b_r \le C - \mathrm{base}(\tau)$.
It is solved exactly by dynamic programming over a discretized capacity,
with weights rounded up so that the result stays feasible. The knapsack
is needed because region types compete for one step-wide capacity.

In the experiments of \S\ref{sec:eval}, blocks of one type may also
select different frontier points, with the sums taken over blocks
(Algorithm~\ref{alg:step} shows the form with one point per type); this avoids an
all-or-nothing choice when a valuable candidate fits on only some
blocks. Larger save sets go to blocks later in forward order: their
backward runs first, so their values are freed earliest and hold memory
for the shortest time. The result lists, for each block, the coordinates
(Eq.~\ref{eq:coord}) of the values its save set saves.

\subsubsection{Shared plans and token buckets}
\label{sec:planning-keys}

On real data, the token count changes from step to step. Solving a new
plan every step would cost solver time and, because every plan changes
the sizes the allocator serves, fragment its memory. The planner
therefore groups token counts into buckets and solves one plan per
bucket, which later steps in the same bucket reuse. Buckets are
geometric: neighboring buckets differ by a factor of 1.1 for primary
regions and of 1.25 for the vision and audio regions, whose token counts
vary more; for the latter, the bucket is that of the largest count seen
so far, so that their swings do not change the plan. We write
$q = \mathrm{bucket}(\tau)$ for the bucket of token count $\tau$ and
$\tau_q^{+}$ for its upper edge. More tokens need more memory, so the
plan of a bucket is solved at its upper edge: each rank's must-only peak
is charged at $\tau_q^{+}$, and the plan fits every step in the bucket.

All ranks run the same plan, because the step time is set by the slowest
rank. At the start of each step, the ranks all-reduce their token counts
(taking the maximum), their remaining capacity (taking the minimum), and
whether any rank needs a new reference or memory sample. Every rank thus
computes the same bucket and finds the same cached plan, or misses
together; on a miss, rank~0 solves for the largest token count and the
smallest capacity and broadcasts the plan.

\section{Experiments}
\label{sec:eval}

We evaluate \sys on three models (Table~\ref{tab:models}). Qwen3.8-27B, a
dense model, and Qwen3.6-35B-A3B, an MoE model, each trained on 16 GPUs,
carry all main experiments; Qwen3.5-397B, trained on 64 GPUs, tests whether
the results hold at scale. The evaluation answers the following questions.
Does \sys reproduce the results of plain execution exactly
(\S\ref{sec:eval-exact})? How much does replay alone gain with the save set
held fixed (\S\ref{sec:eval-replay})? Does planning under a memory limit
outperform full checkpointing and hand-tuned policies, also when the
configuration changes (\S\ref{sec:eval-frontier})? Which design elements
matter, and how does \sys compare with the memory budget of
\texttt{torch.compile} (\S\ref{sec:eval-ablation},
\S\ref{sec:eval-compile})? Do the results hold at scale and on real data,
and what does \sys cost (\S\ref{sec:eval-397b},
\S\ref{sec:eval-overhead})?

\subsection{Experimental Setup}
\label{sec:eval-setup}

Table~\ref{tab:models} lists the models. Each configuration we compare on
them is defined by two choices: the \emph{save set}, which tensors are
saved for the backward pass, and the \emph{recomputation mode}, whether
PyTorch recomputes everything else block by block or \sys replays it.
Table~\ref{tab:configs} arranges the configurations along these two
choices. Comparing the two columns of a row isolates replay
(\S\ref{sec:eval-replay}); comparing rows isolates the choice of tensors,
that is, planning (\S\ref{sec:eval-frontier}).

A hand-tuned save set consists of the outputs of a few operators chosen
from an offline profile, ranked by the recomputation time they remove per
GB saved. H1, chosen for Qwen3.5-397B and adapted to the width of each model,
saves the router's softmax and top-$k$, the GDN gate projections, the
concatenation after RoPE and the output projections; H2 saves the
attention outputs, including those of the vision tower; and H3 saves the
attention outputs and the concatenation after RoPE, and on the 35B model
also the router's top-$k$, an output projection and the GDN scan. Full checkpointing
runs \texttt{torch.\allowbreak utils.\allowbreak checkpoint} with the stock
kernels and hand-tuned SAC runs PyTorch's SAC with the same registered
kernels as \sys; must-only saves only the
mandatory set of \S\ref{sec:tape-pack}, which is empty on our models.

We report the median step time over the steps of a run (the mean on
Qwen3.5-397B), unless noted otherwise, and the peak memory reserved by PyTorch's caching allocator on
rank~0, as a percentage of device memory. The memory limit $f$ is also a
fraction of device memory, and \sys runs with the bucket-edge charge and
per-block points of \S\ref{sec:planning-keys} and
\S\ref{sec:planning-solve}.

\begin{table}[t]
\centering
\footnotesize
\caption{Model architectures and training configurations. All models use FSDP2 across the listed GPUs without additional parallelism.}
\label{tab:models}
\begin{tabular}{@{}lccc@{}}
\toprule
   \textbf{Model}                   & \textbf{Qwen3.8-27B} & \textbf{Qwen3.6-35B-A3B} & \textbf{Qwen3.5-397B} \\
\midrule
Feed-forward          & dense       & MoE 256, top-8  & MoE 512, top-10 \\
GDN blocks            & 48          & 30              & 45 \\
Full-attention blocks & 16          & 10              & 15 \\
\midrule
GPUs (nodes)          & 16 (2)      & 16 (2)          & 64 (8) \\
Local batch size      & 18          & 16, 48          & 16 \\
\bottomrule
\end{tabular}
\end{table}

\begin{table}[t]
\centering
\footnotesize
\caption{Evaluated configurations by save set and recomputation mechanism.}
\label{tab:configs}
\begin{tabular}{@{}lll@{}}
\toprule
\textbf{Save set} & \textbf{Block recomputation} & \textbf{Replay} \\
\midrule
Block inputs only & full checkpointing & must-only ($f = 0$) \\
H1--H3 & hand-tuned SAC & replayed H1--H3 \\
Planned under limit $f$ & -- & \sys at $f$ \\
\bottomrule
\end{tabular}
\end{table}

\subsection{Correctness}
\label{sec:eval-exact}

\begin{figure}[h]
\centering
\subcaptionbox{Moving average over 20 steps}{\begin{tikzpicture}
\begin{axis}[
  width=0.6\linewidth, height=4.3cm,
  xlabel={step}, ylabel={training loss},
  xmin=0, xmax=500, ymin=0.63, ymax=0.93,
  tick label style={font=\scriptsize}, label style={font=\scriptsize},
  yticklabel style={/pgf/number format/fixed,/pgf/number format/precision=2},
  legend style={font=\scriptsize, at={(0.97,0.97)}, anchor=north east,
    draw=none, fill=white, fill opacity=0.8, text opacity=1},
]
\addplot[tsFull, thin] coordinates {(1,0.8541) (2,0.8933) (3,0.9117) (4,0.8947) (5,0.8879) (6,0.8869) (7,0.8826) (8,0.8793) (9,0.8850) (10,0.8826) (11,0.8764) (12,0.8853) (13,0.8890) (14,0.8864) (15,0.8809) (16,0.8760) (17,0.8834) (18,0.8772) (19,0.8751) (20,0.8720) (21,0.8808) (22,0.8805) (23,0.8831) (24,0.8837) (25,0.8789) (26,0.8922) (27,0.8935) (28,0.8953) (29,0.8871) (30,0.8911) (31,0.8952) (32,0.8893) (33,0.8851) (34,0.8825) (35,0.8903) (36,0.8897) (37,0.8809) (38,0.8857) (39,0.8932) (40,0.8995) (41,0.8920) (42,0.8881) (43,0.8791) (44,0.8732) (45,0.8829) (46,0.8665) (47,0.8687) (48,0.8635) (49,0.8713) (50,0.8690) (51,0.8678) (52,0.8589) (53,0.8603) (54,0.8578) (55,0.8548) (56,0.8568) (57,0.8574) (58,0.8576) (59,0.8614) (60,0.8538) (61,0.8524) (62,0.8485) (63,0.8571) (64,0.8669) (65,0.8544) (66,0.8544) (67,0.8459) (68,0.8531) (69,0.8465) (70,0.8463) (71,0.8420) (72,0.8462) (73,0.8427) (74,0.8462) (75,0.8481) (76,0.8446) (77,0.8451) (78,0.8383) (79,0.8268) (80,0.8288) (81,0.8265) (82,0.8285) (83,0.8217) (84,0.8211) (85,0.8272) (86,0.8258) (87,0.8248) (88,0.8177) (89,0.8201) (90,0.8127) (91,0.8116) (92,0.8146) (93,0.8091) (94,0.8086) (95,0.7968) (96,0.8055) (97,0.8120) (98,0.8165) (99,0.8132) (100,0.8063) (101,0.8057) (102,0.8017) (103,0.8010) (104,0.7969) (105,0.7887) (106,0.7836) (107,0.7894) (108,0.7943) (109,0.7909) (110,0.7959) (111,0.7966) (112,0.7953) (113,0.8024) (114,0.8022) (115,0.8069) (116,0.7951) (117,0.7848) (118,0.7827) (119,0.7811) (120,0.7820) (121,0.7763) (122,0.7731) (123,0.7675) (124,0.7598) (125,0.7577) (126,0.7590) (127,0.7510) (128,0.7458) (129,0.7422) (130,0.7392) (131,0.7375) (132,0.7339) (133,0.7217) (134,0.7170) (135,0.7107) (136,0.7178) (137,0.7166) (138,0.7129) (139,0.7150) (140,0.7165) (141,0.7179) (142,0.7203) (143,0.7176) (144,0.7216) (145,0.7250) (146,0.7250) (147,0.7315) (148,0.7332) (149,0.7331) (150,0.7330) (151,0.7319) (152,0.7299) (153,0.7369) (154,0.7345) (155,0.7343) (156,0.7265) (157,0.7288) (158,0.7329) (159,0.7274) (160,0.7286) (161,0.7286) (162,0.7251) (163,0.7283) (164,0.7294) (165,0.7301) (166,0.7324) (167,0.7238) (168,0.7155) (169,0.7162) (170,0.7110) (171,0.7117) (172,0.7127) (173,0.7100) (174,0.7098) (175,0.7177) (176,0.7211) (177,0.7167) (178,0.7120) (179,0.7139) (180,0.7116) (181,0.7111) (182,0.7139) (183,0.7138) (184,0.7110) (185,0.7084) (186,0.7049) (187,0.7065) (188,0.7057) (189,0.7038) (190,0.7010) (191,0.6994) (192,0.6961) (193,0.6932) (194,0.6985) (195,0.6942) (196,0.6912) (197,0.6924) (198,0.6928) (199,0.6924) (200,0.6933) (201,0.6929) (202,0.6895) (203,0.6802) (204,0.6779) (205,0.6876) (206,0.6894) (207,0.6846) (208,0.6905) (209,0.6861) (210,0.6918) (211,0.6927) (212,0.6977) (213,0.6981) (214,0.6942) (215,0.6951) (216,0.6970) (217,0.6960) (218,0.6964) (219,0.6952) (220,0.6954) (221,0.6991) (222,0.7027) (223,0.7089) (224,0.7075) (225,0.6999) (226,0.6939) (227,0.6990) (228,0.6961) (229,0.7024) (230,0.7003) (231,0.6986) (232,0.6994) (233,0.7024) (234,0.6996) (235,0.6958) (236,0.6984) (237,0.6939) (238,0.6939) (239,0.6966) (240,0.6924) (241,0.6918) (242,0.6932) (243,0.6924) (244,0.6980) (245,0.6941) (246,0.6964) (247,0.6937) (248,0.6927) (249,0.6936) (250,0.6953) (251,0.6993) (252,0.6911) (253,0.6884) (254,0.6878) (255,0.6821) (256,0.6808) (257,0.6824) (258,0.6774) (259,0.6767) (260,0.6822) (261,0.6799) (262,0.6820) (263,0.6884) (264,0.6793) (265,0.6837) (266,0.6853) (267,0.6901) (268,0.6930) (269,0.6920) (270,0.6927) (271,0.6903) (272,0.6908) (273,0.6950) (274,0.6963) (275,0.7051) (276,0.6998) (277,0.6983) (278,0.7000) (279,0.6959) (280,0.6956) (281,0.6955) (282,0.6912) (283,0.6854) (284,0.6912) (285,0.6896) (286,0.6904) (287,0.6910) (288,0.6916) (289,0.6890) (290,0.6867) (291,0.6840) (292,0.6857) (293,0.6815) (294,0.6792) (295,0.6750) (296,0.6737) (297,0.6721) (298,0.6772) (299,0.6741) (300,0.6699) (301,0.6668) (302,0.6685) (303,0.6652) (304,0.6625) (305,0.6695) (306,0.6663) (307,0.6613) (308,0.6578) (309,0.6523) (310,0.6534) (311,0.6534) (312,0.6528) (313,0.6510) (314,0.6518) (315,0.6525) (316,0.6575) (317,0.6619) (318,0.6560) (319,0.6615) (320,0.6577) (321,0.6585) (322,0.6598) (323,0.6620) (324,0.6687) (325,0.6615) (326,0.6614) (327,0.6657) (328,0.6672) (329,0.6768) (330,0.6777) (331,0.6774) (332,0.6814) (333,0.6857) (334,0.6891) (335,0.6876) (336,0.6849) (337,0.6858) (338,0.6873) (339,0.6839) (340,0.6859) (341,0.6838) (342,0.6814) (343,0.6836) (344,0.6775) (345,0.6814) (346,0.6861) (347,0.6854) (348,0.6846) (349,0.6788) (350,0.6805) (351,0.6788) (352,0.6798) (353,0.6750) (354,0.6710) (355,0.6716) (356,0.6721) (357,0.6675) (358,0.6700) (359,0.6732) (360,0.6821) (361,0.6825) (362,0.6815) (363,0.6767) (364,0.6768) (365,0.6742) (366,0.6714) (367,0.6670) (368,0.6653) (369,0.6684) (370,0.6629) (371,0.6679) (372,0.6665) (373,0.6706) (374,0.6736) (375,0.6744) (376,0.6762) (377,0.6757) (378,0.6724) (379,0.6693) (380,0.6649) (381,0.6649) (382,0.6660) (383,0.6736) (384,0.6726) (385,0.6750) (386,0.6731) (387,0.6750) (388,0.6796) (389,0.6788) (390,0.6828) (391,0.6767) (392,0.6757) (393,0.6731) (394,0.6752) (395,0.6730) (396,0.6716) (397,0.6733) (398,0.6758) (399,0.6762) (400,0.6709) (401,0.6711) (402,0.6710) (403,0.6659) (404,0.6629) (405,0.6565) (406,0.6557) (407,0.6541) (408,0.6497) (409,0.6476) (410,0.6450) (411,0.6462) (412,0.6479) (413,0.6502) (414,0.6491) (415,0.6504) (416,0.6463) (417,0.6446) (418,0.6448) (419,0.6487) (420,0.6519) (421,0.6598) (422,0.6577) (423,0.6556) (424,0.6582) (425,0.6578) (426,0.6606) (427,0.6590) (428,0.6584) (429,0.6589) (430,0.6576) (431,0.6642) (432,0.6633) (433,0.6644) (434,0.6624) (435,0.6644) (436,0.6742) (437,0.6771) (438,0.6774) (439,0.6809) (440,0.6797) (441,0.6763) (442,0.6735) (443,0.6710) (444,0.6718) (445,0.6723) (446,0.6680) (447,0.6695) (448,0.6745) (449,0.6784) (450,0.6790) (451,0.6771) (452,0.6728) (453,0.6745) (454,0.6749) (455,0.6735) (456,0.6706) (457,0.6721) (458,0.6700) (459,0.6650) (460,0.6661) (461,0.6645) (462,0.6589) (463,0.6640) (464,0.6646) (465,0.6681) (466,0.6686) (467,0.6677) (468,0.6645) (469,0.6660) (470,0.6612) (471,0.6604) (472,0.6637) (473,0.6613) (474,0.6658) (475,0.6635) (476,0.6601) (477,0.6638) (478,0.6636) (479,0.6636) (480,0.6678) (481,0.6677) (482,0.6758) (483,0.6776) (484,0.6788) (485,0.6756) (486,0.6814) (487,0.6804) (488,0.6771) (489,0.6702) (490,0.6751) (491,0.6769) (492,0.6751) (493,0.6728) (494,0.6682) (495,0.6688) (496,0.6694) (497,0.6633) (498,0.6644) (499,0.6646) (500,0.6618)};
\addlegendentry{full checkpointing}
\addplot[tsOurs, thin, densely dashed] coordinates {(1,0.8541) (2,0.8933) (3,0.9116) (4,0.8945) (5,0.8877) (6,0.8869) (7,0.8826) (8,0.8793) (9,0.8851) (10,0.8827) (11,0.8765) (12,0.8853) (13,0.8891) (14,0.8864) (15,0.8810) (16,0.8760) (17,0.8834) (18,0.8772) (19,0.8751) (20,0.8721) (21,0.8809) (22,0.8806) (23,0.8832) (24,0.8838) (25,0.8791) (26,0.8923) (27,0.8937) (28,0.8954) (29,0.8873) (30,0.8912) (31,0.8953) (32,0.8894) (33,0.8852) (34,0.8825) (35,0.8903) (36,0.8897) (37,0.8810) (38,0.8858) (39,0.8933) (40,0.8996) (41,0.8920) (42,0.8881) (43,0.8791) (44,0.8731) (45,0.8829) (46,0.8664) (47,0.8686) (48,0.8634) (49,0.8712) (50,0.8690) (51,0.8677) (52,0.8588) (53,0.8602) (54,0.8577) (55,0.8547) (56,0.8567) (57,0.8572) (58,0.8575) (59,0.8612) (60,0.8536) (61,0.8522) (62,0.8483) (63,0.8569) (64,0.8667) (65,0.8542) (66,0.8543) (67,0.8458) (68,0.8529) (69,0.8464) (70,0.8461) (71,0.8418) (72,0.8460) (73,0.8425) (74,0.8460) (75,0.8479) (76,0.8444) (77,0.8449) (78,0.8380) (79,0.8265) (80,0.8286) (81,0.8263) (82,0.8283) (83,0.8214) (84,0.8208) (85,0.8269) (86,0.8255) (87,0.8245) (88,0.8175) (89,0.8197) (90,0.8124) (91,0.8113) (92,0.8143) (93,0.8089) (94,0.8083) (95,0.7966) (96,0.8052) (97,0.8117) (98,0.8163) (99,0.8131) (100,0.8062) (101,0.8056) (102,0.8015) (103,0.8008) (104,0.7967) (105,0.7886) (106,0.7835) (107,0.7892) (108,0.7941) (109,0.7907) (110,0.7957) (111,0.7964) (112,0.7950) (113,0.8021) (114,0.8019) (115,0.8066) (116,0.7948) (117,0.7844) (118,0.7822) (119,0.7807) (120,0.7816) (121,0.7759) (122,0.7727) (123,0.7671) (124,0.7595) (125,0.7573) (126,0.7586) (127,0.7507) (128,0.7455) (129,0.7419) (130,0.7389) (131,0.7372) (132,0.7336) (133,0.7215) (134,0.7168) (135,0.7105) (136,0.7176) (137,0.7164) (138,0.7127) (139,0.7149) (140,0.7163) (141,0.7178) (142,0.7202) (143,0.7175) (144,0.7215) (145,0.7248) (146,0.7249) (147,0.7313) (148,0.7330) (149,0.7329) (150,0.7328) (151,0.7318) (152,0.7298) (153,0.7368) (154,0.7344) (155,0.7342) (156,0.7265) (157,0.7287) (158,0.7328) (159,0.7272) (160,0.7284) (161,0.7285) (162,0.7250) (163,0.7281) (164,0.7293) (165,0.7300) (166,0.7323) (167,0.7237) (168,0.7154) (169,0.7162) (170,0.7109) (171,0.7116) (172,0.7126) (173,0.7099) (174,0.7097) (175,0.7176) (176,0.7210) (177,0.7166) (178,0.7119) (179,0.7139) (180,0.7116) (181,0.7111) (182,0.7139) (183,0.7138) (184,0.7110) (185,0.7084) (186,0.7049) (187,0.7064) (188,0.7056) (189,0.7037) (190,0.7009) (191,0.6992) (192,0.6960) (193,0.6931) (194,0.6983) (195,0.6940) (196,0.6910) (197,0.6921) (198,0.6925) (199,0.6921) (200,0.6931) (201,0.6926) (202,0.6893) (203,0.6800) (204,0.6777) (205,0.6874) (206,0.6892) (207,0.6845) (208,0.6904) (209,0.6860) (210,0.6917) (211,0.6926) (212,0.6977) (213,0.6981) (214,0.6942) (215,0.6951) (216,0.6970) (217,0.6960) (218,0.6964) (219,0.6952) (220,0.6953) (221,0.6991) (222,0.7026) (223,0.7089) (224,0.7075) (225,0.6998) (226,0.6938) (227,0.6989) (228,0.6960) (229,0.7023) (230,0.7003) (231,0.6986) (232,0.6994) (233,0.7024) (234,0.6995) (235,0.6958) (236,0.6985) (237,0.6939) (238,0.6940) (239,0.6966) (240,0.6925) (241,0.6920) (242,0.6933) (243,0.6926) (244,0.6982) (245,0.6944) (246,0.6966) (247,0.6940) (248,0.6931) (249,0.6939) (250,0.6956) (251,0.6996) (252,0.6915) (253,0.6888) (254,0.6883) (255,0.6826) (256,0.6812) (257,0.6829) (258,0.6777) (259,0.6771) (260,0.6825) (261,0.6802) (262,0.6823) (263,0.6887) (264,0.6796) (265,0.6839) (266,0.6855) (267,0.6902) (268,0.6931) (269,0.6921) (270,0.6928) (271,0.6903) (272,0.6908) (273,0.6950) (274,0.6963) (275,0.7051) (276,0.6998) (277,0.6983) (278,0.7001) (279,0.6959) (280,0.6956) (281,0.6956) (282,0.6913) (283,0.6855) (284,0.6913) (285,0.6898) (286,0.6905) (287,0.6912) (288,0.6918) (289,0.6892) (290,0.6869) (291,0.6842) (292,0.6859) (293,0.6817) (294,0.6794) (295,0.6751) (296,0.6738) (297,0.6723) (298,0.6774) (299,0.6742) (300,0.6701) (301,0.6670) (302,0.6687) (303,0.6654) (304,0.6625) (305,0.6695) (306,0.6663) (307,0.6613) (308,0.6578) (309,0.6522) (310,0.6533) (311,0.6533) (312,0.6526) (313,0.6509) (314,0.6517) (315,0.6524) (316,0.6573) (317,0.6617) (318,0.6559) (319,0.6614) (320,0.6576) (321,0.6583) (322,0.6597) (323,0.6619) (324,0.6688) (325,0.6616) (326,0.6616) (327,0.6658) (328,0.6673) (329,0.6769) (330,0.6777) (331,0.6773) (332,0.6814) (333,0.6858) (334,0.6891) (335,0.6877) (336,0.6850) (337,0.6858) (338,0.6875) (339,0.6840) (340,0.6860) (341,0.6838) (342,0.6815) (343,0.6837) (344,0.6776) (345,0.6814) (346,0.6861) (347,0.6854) (348,0.6845) (349,0.6787) (350,0.6804) (351,0.6788) (352,0.6799) (353,0.6750) (354,0.6711) (355,0.6717) (356,0.6722) (357,0.6677) (358,0.6701) (359,0.6732) (360,0.6821) (361,0.6825) (362,0.6814) (363,0.6766) (364,0.6766) (365,0.6740) (366,0.6712) (367,0.6669) (368,0.6652) (369,0.6684) (370,0.6629) (371,0.6679) (372,0.6664) (373,0.6705) (374,0.6735) (375,0.6743) (376,0.6761) (377,0.6755) (378,0.6723) (379,0.6692) (380,0.6649) (381,0.6649) (382,0.6659) (383,0.6736) (384,0.6726) (385,0.6750) (386,0.6731) (387,0.6750) (388,0.6796) (389,0.6787) (390,0.6827) (391,0.6766) (392,0.6754) (393,0.6727) (394,0.6748) (395,0.6725) (396,0.6710) (397,0.6727) (398,0.6752) (399,0.6755) (400,0.6701) (401,0.6704) (402,0.6702) (403,0.6651) (404,0.6621) (405,0.6557) (406,0.6550) (407,0.6534) (408,0.6489) (409,0.6468) (410,0.6443) (411,0.6455) (412,0.6473) (413,0.6496) (414,0.6486) (415,0.6500) (416,0.6459) (417,0.6443) (418,0.6445) (419,0.6484) (420,0.6516) (421,0.6595) (422,0.6573) (423,0.6553) (424,0.6578) (425,0.6574) (426,0.6602) (427,0.6585) (428,0.6580) (429,0.6584) (430,0.6572) (431,0.6638) (432,0.6630) (433,0.6640) (434,0.6620) (435,0.6639) (436,0.6736) (437,0.6765) (438,0.6769) (439,0.6804) (440,0.6793) (441,0.6760) (442,0.6732) (443,0.6708) (444,0.6716) (445,0.6721) (446,0.6678) (447,0.6693) (448,0.6742) (449,0.6781) (450,0.6787) (451,0.6766) (452,0.6723) (453,0.6739) (454,0.6743) (455,0.6729) (456,0.6702) (457,0.6716) (458,0.6696) (459,0.6645) (460,0.6656) (461,0.6639) (462,0.6583) (463,0.6633) (464,0.6639) (465,0.6674) (466,0.6678) (467,0.6670) (468,0.6638) (469,0.6653) (470,0.6605) (471,0.6597) (472,0.6630) (473,0.6606) (474,0.6651) (475,0.6628) (476,0.6594) (477,0.6632) (478,0.6629) (479,0.6628) (480,0.6670) (481,0.6668) (482,0.6750) (483,0.6768) (484,0.6780) (485,0.6748) (486,0.6805) (487,0.6795) (488,0.6761) (489,0.6692) (490,0.6741) (491,0.6759) (492,0.6741) (493,0.6717) (494,0.6671) (495,0.6677) (496,0.6682) (497,0.6621) (498,0.6632) (499,0.6633) (500,0.6606)};
\addlegendentry{\sys}
\end{axis}
\end{tikzpicture}}\hfill
\subcaptionbox{Steps 300--340}{\begin{tikzpicture}
\begin{axis}[
  width=0.43\linewidth, height=4.3cm, ytick pos=right, yticklabel pos=right,
  xlabel={step}, xmin=298, xmax=342, ymin=0.52, ymax=0.84,
  tick label style={font=\scriptsize}, label style={font=\scriptsize},
  yticklabel style={/pgf/number format/fixed,/pgf/number format/precision=1},
]
\addplot[tsFull, thin, mark=*, mark size=0.7pt] coordinates {(300,0.6719) (301,0.6478) (302,0.7199) (303,0.6075) (304,0.6483) (305,0.8140) (306,0.6380) (307,0.6458) (308,0.6732) (309,0.5505) (310,0.7159) (311,0.6608) (312,0.6330) (313,0.6057) (314,0.6141) (315,0.6718) (316,0.7245) (317,0.6763) (318,0.6314) (319,0.6799) (320,0.5962) (321,0.6621) (322,0.7461) (323,0.6518) (324,0.7826) (325,0.6705) (326,0.6365) (327,0.7309) (328,0.7034) (329,0.7414) (330,0.7344) (331,0.6544) (332,0.7133) (333,0.6923) (334,0.6810) (335,0.6433) (336,0.6694) (337,0.6939) (338,0.6628) (339,0.6105) (340,0.6374)};
\addplot[tsOurs, thin, densely dashed, mark=o, mark size=1.3pt] coordinates {(300,0.6729) (301,0.6478) (302,0.7196) (303,0.6075) (304,0.6464) (305,0.8133) (306,0.6385) (307,0.6456) (308,0.6729) (309,0.5500) (310,0.7165) (311,0.6603) (312,0.6327) (313,0.6049) (314,0.6144) (315,0.6718) (316,0.7240) (317,0.6770) (318,0.6311) (319,0.6802) (320,0.5970) (321,0.6628) (322,0.7462) (323,0.6526) (324,0.7835) (325,0.6706) (326,0.6371) (327,0.7309) (328,0.7028) (329,0.7408) (330,0.7339) (331,0.6527) (332,0.7143) (333,0.6930) (334,0.6805) (335,0.6434) (336,0.6691) (337,0.6944) (338,0.6637) (339,0.6102) (340,0.6377)};
\end{axis}
\end{tikzpicture}}
\caption{Training-loss trajectories for full checkpointing and \sys on the same real-data sequence (\S\ref{sec:eval-realdata}).}
\label{fig:loss}
\end{figure}
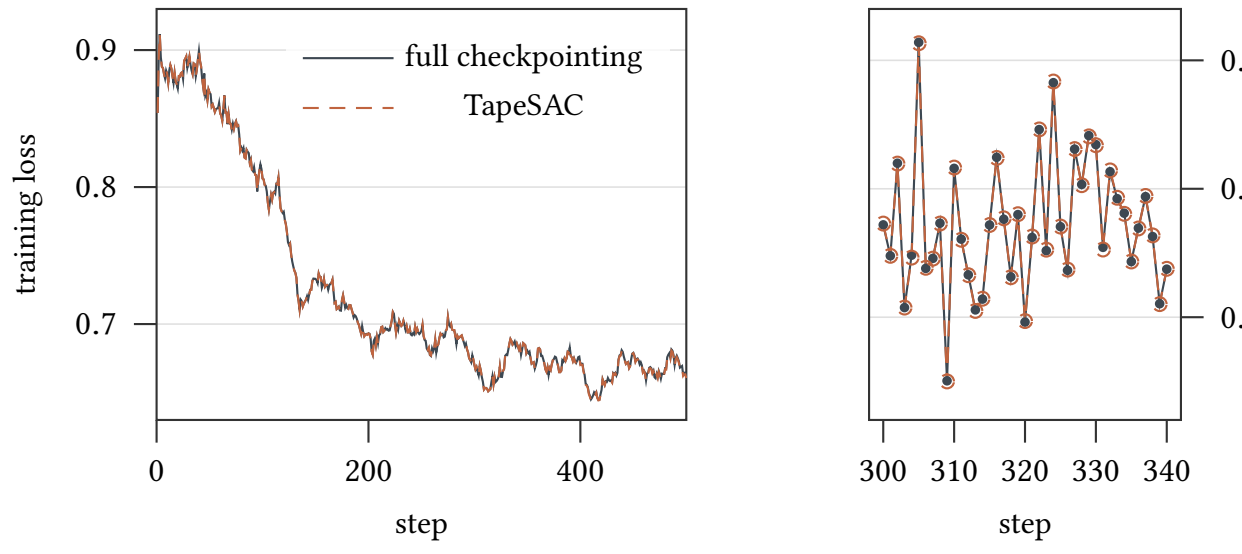

Because replay applies the same operators to the same inputs as the
forward pass, it should reproduce every recomputed value bitwise,
independently of the save set; likewise,
registering a kernel as a custom operator should leave its results
unchanged (\S\ref{sec:visibility}). We verify both properties at three
levels of granularity. At the operator level, every registered kernel
agrees with its stock implementation in outputs and gradients and passes
\texttt{torch.library.opcheck}. At the block level, a GDN block and a
full-attention block of Qwen3.5-397B, replayed without saving any planned
value so that every node that backward needs is recomputed, reproduce
the outputs and all parameter gradients of plain execution bitwise. At
the model level, the first training step of each of the three models
yields the same loss under full checkpointing, every hand-tuned policy in
either execution mode, and every \sys plan we evaluate, up to one unit in
the last place of float32, a difference that also occurs between
PyTorch's own configurations. Over 500 steps of real data
(\S\ref{sec:eval-realdata}), the training loss of \sys follows that of
full checkpointing step by step (Figure~\ref{fig:loss}).

A suite of unit tests on a small network with views, transposes and
in-place updates further verifies the mechanisms of \S\ref{sec:tape}:
replay skips dead branches, executes fewer nodes than were recorded,
and leaves saved values intact; random operators and mutations of
leaves are rejected; and a module that computes indices with
\texttt{nonzero} and \texttt{repeat\_interleave} obtains bitwise
identical gradients without host synchronization in the backward pass,
since the outputs of these operators belong to the mandatory set and are
saved rather than replayed.

\subsection{Replay at a Fixed Save Set}
\label{sec:eval-replay}

\begin{figure}[h]
\centering
\begin{tikzpicture}
\begin{axis}[
  width=\columnwidth, height=4.3cm,
  ybar, bar width=10pt,
  legend image code/.code={\draw[#1] (0cm,-0.08cm) rectangle (0.2cm,0.12cm);},
  xmin=0.3, xmax=12.0, ymin=0, ymax=3.8,
  ytick={0,1,2,3}, ylabel={step-time reduction (\%)},
  xtick={2.5,7.1,10.7}, major x tick style={draw=none},
  xticklabels={27B,35B-A3B,397B},
  xticklabel style={align=center},
  tick label style={font=\scriptsize}, label style={font=\footnotesize},
  nodes near coords, point meta=explicit symbolic,
  every node near coord/.append style={font=\scriptsize, inner sep=1.5pt},
  legend style={font=\scriptsize, draw=none, fill=none, at={(0.99,0.97)},
    anchor=north east, legend columns=4,
    /tikz/every even column/.append style={column sep=5pt}},
]
\addplot[bar shift=0pt, fill=tsFull, draw=tsFull!70!black] coordinates
  {(1,3.021) [3.0] (5.6,1.710) [1.7] (10.2,1.781) [1.8]};
\addplot[bar shift=0pt, fill=tsBase, draw=tsBase!70!black] coordinates
  {(2,2.704) [2.7] (6.6,1.304) [1.3] (11.2,0.174) [0.2]};
\addplot[bar shift=0pt, fill=tsBaseMid, draw=tsBaseMid!70!black] coordinates
  {(3,0.031) [0.0] (7.6,0.914) [0.9]};
\addplot[bar shift=0pt, fill=tsBaseLight, draw=tsBaseLight!60!black] coordinates
  {(4,0.112) [0.1] (8.6,0.954) [1.0]};
\legend{full checkpointing, H1, H2, H3}
\end{axis}
\end{tikzpicture}
\caption{Step-time reduction from tensor-level replay relative to block recomputation for each fixed save set.}
\label{fig:replay}
\end{figure}
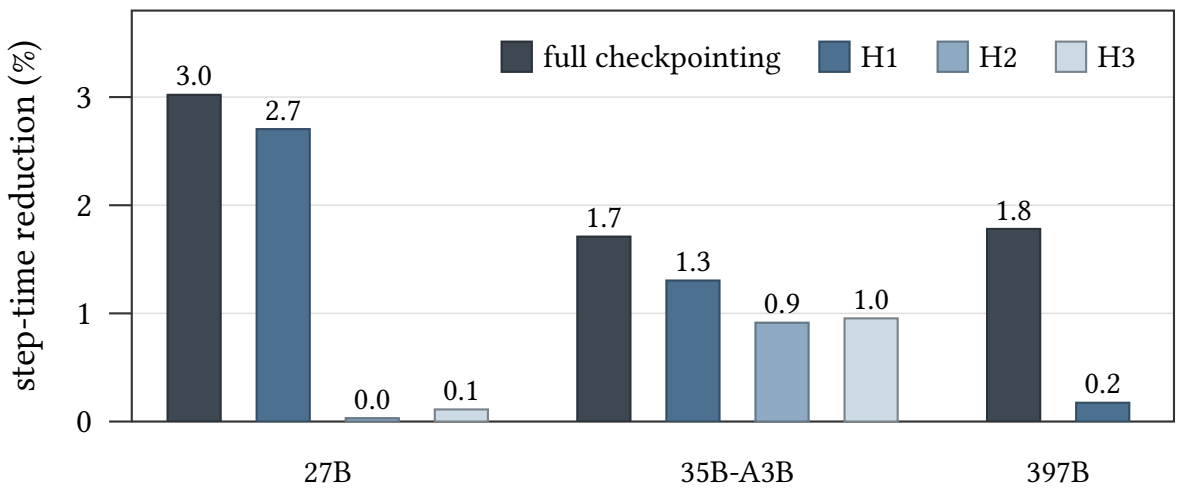

\begin{figure*}[t]
\centering
\subcaptionbox{Qwen3.8-27B}{\begin{tikzpicture}
\begin{axis}[
  width=0.49\textwidth, height=4.8cm,
  xlabel={reserved memory, rank 0 (\% of device)}, ylabel={step time (s)},
  xmin=62.0, xmax=100.1, ymin=29.1, ymax=31.45,
  tick label style={font=\scriptsize}, label style={font=\footnotesize},
  yticklabel style={/pgf/number format/fixed,/pgf/number format/precision=1},
]
\addplot[tsOurs, thick, mark=*, mark size=1.4pt] coordinates
  {(64.4,30.306) (74.5,29.639) (75.6,29.552) (78.1,29.468) (80.5,29.401) (82.9,29.296) (84.6,29.222) (86.9,29.154)};
\addplot[only marks, mark=square*, mark size=1.8pt, tsFull] coordinates {(65.4,31.250)};
\addplot[only marks, mark=diamond*, mark size=2.4pt, tsBase] coordinates {(97.6,30.626) (86.2,29.430) (82.5,29.502)};
\addplot[only marks, mark=diamond, mark size=2.4pt, tsBase, thick] coordinates {(97.9,29.798) (86.2,29.421) (81.6,29.469)};
\node[font=\scriptsize, tsFull, anchor=west] at (axis cs:65.9,31.250) {full checkpointing};
\node[font=\scriptsize, tsOurs, anchor=south west] at (axis cs:64.9,30.33) {must-only};
\node[font=\scriptsize, tsBase, anchor=east] at (axis cs:96.9,30.626) {H1};
\node[font=\scriptsize, tsBase, anchor=east] at (axis cs:97.1,29.80) {H1 rp};
\node[font=\scriptsize, tsBase, anchor=south] at (axis cs:86.3,29.47) {H2};
\node[font=\scriptsize, tsBase, anchor=south] at (axis cs:81.8,29.53) {H3};
\node[font=\scriptsize, tsOurs, anchor=north east] at (axis cs:74.4,29.62) {0.78};
\node[font=\scriptsize, tsOurs, anchor=north east] at (axis cs:82.9,29.27) {0.90};
\end{axis}
\end{tikzpicture}}\hfill
\subcaptionbox{Qwen3.6-35B-A3B}{\begin{tikzpicture}
\begin{axis}[
  width=0.49\textwidth, height=4.8cm,
  xlabel={reserved memory, rank 0 (\% of device)}, ylabel={step time (s)},
  xmin=28.0, xmax=78.4, ymin=9.0, ymax=10.7,
  tick label style={font=\scriptsize}, label style={font=\footnotesize},
  yticklabel style={/pgf/number format/fixed,/pgf/number format/precision=1},
]
\addplot[tsOurs, thick, mark=*, mark size=1.4pt] coordinates
  {(31.0,10.402) (38.1,9.960) (42.2,9.807) (50.5,9.539) (58.5,9.347) (67.0,9.216) (74.9,9.115)};
\addplot[only marks, mark=square*, mark size=1.8pt, tsFull] coordinates {(31.0,10.583)};
\addplot[only marks, mark=diamond*, mark size=2.4pt, tsBase] coordinates {(40.4,10.351) (42.8,10.071) (49.5,10.060)};
\addplot[only marks, mark=diamond, mark size=2.4pt, tsBase, thick] coordinates {(40.4,10.216) (42.8,9.979) (49.5,9.964)};
\node[font=\scriptsize, tsFull, anchor=west] at (axis cs:31.6,10.583) {full checkpointing};
\node[font=\scriptsize, tsOurs, anchor=west] at (axis cs:31.6,10.41) {must-only};
\node[font=\scriptsize, tsBase, anchor=west] at (axis cs:41.0,10.36) {H1};
\node[font=\scriptsize, tsBase, anchor=south] at (axis cs:43.7,10.09) {H2};
\node[font=\scriptsize, tsBase, anchor=south] at (axis cs:49.5,10.08) {H3};
\node[font=\scriptsize, tsOurs, anchor=north] at (axis cs:38.1,9.94) {0.45};
\node[font=\scriptsize, tsOurs, anchor=south] at (axis cs:74.9,9.13) {0.90};
\end{axis}
\end{tikzpicture}}
\caption{Reserved-memory--step-time trade-offs on the two main models. Curves show \sys plans from must-only upward, labeled by memory limit $f$; squares denote full checkpointing, and diamonds denote hand-tuned save sets under PyTorch SAC (filled) or \sys replay (open).}
\label{fig:frontier}
\end{figure*}

Replay addresses the second problem of \S\ref{sec:intro}: block recomputation
reruns a block in program order, running operators whose outputs no
backward node reads and keeping the recomputed values alive together.
To separate replay from the choice of tensors, we hold the save set
fixed and change only how it is recomputed (Figure~\ref{fig:replay}).
We use four save sets: that of full checkpointing, whose replay is
must-only, and the three hand-tuned save sets H1--H3 of
\S\ref{sec:eval-setup}, which save more.

For every save set and model, replay is at least as fast as
block recomputation, and it gains most with the save set of full checkpointing,
which leaves the most to recompute (Figure~\ref{fig:replay}). With that
save set, replay shortens the step by 3.0\% on the 27B model, 1.7\% on the 35B
model and 1.8\% on the 397B model. With the hand-tuned save sets, the gain is
0.9--1.3\% on the 35B model and at most 0.2\% for H2 and H3 on the 27B
model and for H1 on the 397B model. H1 on the 27B model is an outlier at
2.7\%: it nearly fills device memory, and under block recomputation every step
incurs an allocator retry.

\begin{table}[t]
\centering
\footnotesize
\caption{GPU-time breakdown of must-only replay relative to full checkpointing for one profiled step of the 27B model, grouped by the operators launching the kernels.}
\label{tab:breakdown}
\begin{tabular}{@{}lr@{}}
\toprule
\textbf{Item} & \textbf{GPU time (ms)} \\
\midrule
Text MLP down projection, 64 recomputations     & $-1000.6$ \\
Vision MLP output projection, 27 recomputations & $-97.6$ \\
Residual additions, 64 + 27 recomputations      & $-49.1$ \\
\quad Recomputation skipped in total            & $-1147.3$ \\
Extra copies around the attention operator      & $+79.6$ \\
\midrule
Compute stream busy (measured)                  & $-1036$ \\
Step time (measured, profiled step)             & $-973$ \\
\bottomrule
\end{tabular}
\end{table}

Profiling one step of full checkpointing and of must-only on rank 0 of
the 27B model shows where the gain comes from
(Table~\ref{tab:breakdown}). The kernels launched by the forward pass
take the same time in both modes (8.86 and 8.88\,s), so recording adds no
GPU time; the difference lies in the backward pass, and nearly all of it
is one operator, the down projection of each text MLP, a 15.6\,ms GEMM.
Its backward needs its input and its weight but not its output. Autograd
saves that input only when the GEMM executes, so block recomputation must run
the GEMM and discard its result, whereas replay runs only the producers
of the input. The output projection of each vision MLP and the residual
addition at the end of each block are skipped for the same reason, while
attention, the GDN core and the normalizations, whose saved tensors the
backward pass reads, run the same kernels for the same time in both
modes. Of the 1.15\,s of recomputation skipped, 0.97\,s appears as a shorter
step; the rest is absorbed by copies around the attention operator,
which runs FlashAttention-4 as a registered custom operator, and by
longer waits in reduce-scatter on the ranks that compute faster. This is
the redundancy of block recomputation described in \S\ref{sec:bg-eager},
measured on a large model: $T_{\mathrm{rp}}(S) < T_{\mathrm{blk}}(S)$ for
the same $S$ (Eqs.~\ref{eq:t-re} and~\ref{eq:t-rp}).

The 35B model shows the same mechanism: its forward kernels take the
same time in both modes, and its backward kernels take 227\,ms less
under replay, mostly in operators whose outputs the backward pass does
not read, such as the output projections of the vision and audio MLPs
and the down GEMMs of the MoE layers. With H1, they take only 53\,ms
less: a hand-tuned save set already avoids part of the redundant
recomputation and leaves less for replay to skip, which is why replay
gains less with the hand-tuned save sets.

Because replay materializes each recomputed value only when it is
consumed and frees it after its last use, it can also lower memory. With
the save set of full checkpointing, reserved memory is 1.5\% lower on the
27B model and unchanged on the two MoE models, and with H1 it is 2.1\%
lower on the 397B model (Figures~\ref{fig:frontier}
and~\ref{fig:frontier-397b}). The differences are small because the
step's peak is set largely by the sharded parameters, gradients,
communication buffers and fragmentation (\S\ref{sec:eval-memmodel}).
With the save set of full checkpointing, replay thus does not increase
memory:
must-only, the point from which planning starts in
\S\ref{sec:eval-frontier}, is already faster than full checkpointing at
the same or lower memory.

\subsection{Planning under a Memory Limit}
\label{sec:eval-frontier}

Planning trades memory for step time from must-only upward
(Figure~\ref{fig:frontier}). On both main models, more memory gives a
shorter step, with diminishing returns: on the 27B model, the first 10\%
of device memory beyond must-only shortens the step by 0.67\,s,
and the next 10\% by 0.42\,s. How much planning adds depends on the room above full checkpointing. At
$f = 0.90$, \sys is 6.3\% faster than full checkpointing on the 27B
model, where full checkpointing already uses 65\% of device memory, and
13.9\% faster on the 35B model, where it uses only 31\% and the plan
raises reserved memory to 75\%, saving mostly in the vision and
full-attention blocks.

\begin{figure}[h]
\centering
\begin{tikzpicture}
\begin{axis}[
  width=\columnwidth, height=4.2cm,
  ybar stacked, bar width=13pt, clip=false,
  xmin=0.3, xmax=7.7, ymin=0, ymax=6.0,
  xtick={1,2,3,5,6,7}, xticklabels={H1$^{*}$,H2,H3,H1,H2,H3},
  major x tick style={draw=none},
  ytick={0,1,2,3,4,5}, ylabel={step-time reduction (\%)},
  tick label style={font=\scriptsize}, label style={font=\footnotesize},
  legend image code/.code={\draw[#1] (0cm,-0.08cm) rectangle (0.2cm,0.12cm);},
  legend style={font=\scriptsize, draw=none, fill=none, at={(0.27,0.97)},
    anchor=north west, legend columns=2,
    /tikz/every even column/.append style={column sep=5pt}},
]
\addplot[fill=tsOursLight, draw=tsOursLight!70!black] coordinates
  {(1,2.704) (2,0.031) (3,0.112) (5,1.304) (6,0.914) (7,0.954)};
\addplot[fill=tsOurs, draw=tsOurs!70!black] coordinates
  {(1,2.116) (2,0.914) (3,0.505) (5,3.381) (6,1.976) (7,4.115)};
\legend{replay, planning}
\node[font=\scriptsize, anchor=south, inner sep=1.5pt] at (axis cs:1,4.819) {4.8$^{*}$};
\node[font=\scriptsize, anchor=south, inner sep=1.5pt] at (axis cs:2,0.945) {0.9};
\node[font=\scriptsize, anchor=south, inner sep=1.5pt] at (axis cs:3,0.617) {0.6};
\node[font=\scriptsize, anchor=south, inner sep=1.5pt] at (axis cs:5,4.685) {4.7};
\node[font=\scriptsize, anchor=south, inner sep=1.5pt] at (axis cs:6,2.889) {2.9};
\node[font=\scriptsize, anchor=south, inner sep=1.5pt] at (axis cs:7,5.070) {5.1};
\node[font=\scriptsize, anchor=north] at (axis cs:2,0) [yshift=-10pt] {27B};
\node[font=\scriptsize, anchor=north] at (axis cs:6,0) [yshift=-10pt] {35B-A3B};
\end{axis}
\end{tikzpicture}
\caption{Step-time reduction of \sys relative to hand-tuned PyTorch SAC policies at the same or lower reserved memory, decomposed into replay and save-set planning gains. $^{*}$\sys uses 11\% less reserved memory.}
\label{fig:samemem}
\end{figure}
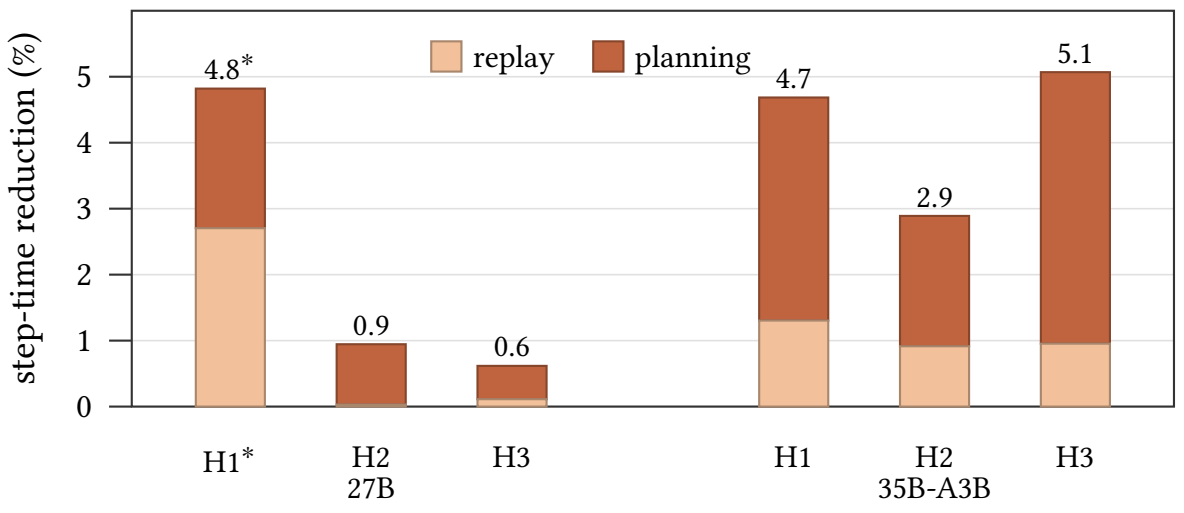

The hand-tuned policies lie off this frontier: at the same reserved
memory, or less for H1 on the 27B model, \sys is faster than every one of
them (Figure~\ref{fig:samemem}). On the 35B model, it is 2.9--5.1\%
faster than the three policies under PyTorch's SAC. Replaying a policy
recovers about 1\% of this gap; the remaining 2.0--4.1\% comes from the
choice of tensors rather than from their execution. On the 27B model, H2
and H3, selected for this model, come within 0.6--0.9\% of \sys, whereas
H1, tuned for a different model, needs 97.6\% of device memory, and
\sys, using 86.9\%, is 4.8\% faster. A hand-tuned policy can thus approach the frontier,
but only for the configuration it was tuned for.

Such a policy fixes which tensors to save, not how much memory to use,
so a change of configuration moves its memory with it. On the 35B
model, the three policies shorten the step by 2.2--4.9\% at a local
batch size of 16, but all of them run out of memory at 48, under
PyTorch's SAC and replayed alike; \sys, with the same $f = 0.90$,
re-plans the save set and remains 5.6\% faster than full checkpointing.

The planner chooses among plans by their predicted speedup, which tracks
the measured one. Relative to must-only, the measured speedup is 89--95\% of the predicted
speedup on the 27B model and 77--79\% on the 35B model. Timing each node
with events in isolation overestimates replay that overlaps other
kernels, but it does so consistently: on both models, the measured order
of the plans matches the predicted one at every limit, which is all the
planner needs to choose among them.

\begin{table}[t]
\centering
\footnotesize
\caption{Effect of solver options on median step time (s) under different memory limits.}
\label{tab:ablation}
\begin{tabular}{@{}lrrrr@{}}
\toprule
\textbf{Variant} & \multicolumn{2}{c}{\textbf{27B}} & \multicolumn{2}{c}{\textbf{35B}} \\
        & \textbf{$f=0.78$} & \textbf{$0.90$} & \textbf{$0.45$} & \textbf{$0.90$} \\
\midrule
both options (default)        & 29.71 & 29.38 & 9.96  & 9.12 \\
neither option                & 30.27 & 29.42 & 10.05 & 9.17 \\
bucket-edge charge only       & 29.70 & 29.42 & 10.07 & 9.14 \\
per-block points only         & 29.82 & 29.46 & 9.94  & 9.13 \\
\bottomrule
\end{tabular}
\end{table}

\label{sec:eval-memmodel}The planner's memory constraint is nearly exact as well. In all three
models, the peak falls at the start of the backward pass, in the window of
the last GDN block, which runs after only the last full-attention block
has freed its saved values; Eq.~\ref{eq:step-peak} therefore exceeds
the peak by at most one block's saved bytes. Most of the gap between
reserved and allocated memory does not depend on what is saved: on the
27B model, 7.8\% of device memory at the peak is idle memory in the
communication pool and in the optimizer's side streams, each of which the
caching allocator serves from its own segment.

\begin{figure}[t]
\centering
\begin{tikzpicture}
\begin{axis}[
  width=\linewidth, height=5.0cm, clip=false,
  xlabel={extra memory (\%)}, ylabel={recomputation eliminated (\%)},
  xmin=-22, xmax=106, ymin=-14, ymax=106,
  xtick={0,20,40,60,80,100}, ytick={0,25,50,75,100},
  tick label style={font=\scriptsize}, label style={font=\footnotesize},
  legend style={font=\scriptsize, at={(0.98,0.04)}, anchor=south east,
    draw=none, fill=white, fill opacity=0.85, text opacity=1},
]
\addplot[tsOurs, thick, mark=*, mark size=1.4pt] coordinates
  {(0,14) (10,43) (21,58) (38,72) (52,82) (65,91) (79,94)};
\addlegendentry{\sys (eager)}
\addplot[tsBase, thick, dashed, mark=square*, mark size=1.4pt] coordinates
  {(0,-4) (3,2) (9,8) (18,24) (29,40) (37,52) (49,69) (49,71) (62,94) (100,100)};
\addlegendentry{memory budget (compiled)}
\addplot[only marks, mark=*, mark size=2pt, tsFull, forget plot] coordinates {(0,0) (100,100)};
\node[font=\scriptsize, tsFull, anchor=east] at (axis cs:-1.5,0) {full ckpt.};
\node[font=\scriptsize, tsFull, anchor=north east] at (axis cs:99,96) {no ckpt.};
\node[font=\scriptsize, tsOurs, anchor=east] at (axis cs:-1.5,14) {must-only};
\node[font=\scriptsize, tsOurs, anchor=south east] at (axis cs:9,44) {0.30};
\node[font=\scriptsize, tsOurs, anchor=north west] at (axis cs:66,90) {0.50};
\node[font=\scriptsize, tsBase, anchor=west] at (axis cs:2.5,-7) {budget 0};
\node[font=\scriptsize, tsBase, anchor=north west] at (axis cs:10,7) {0.1};
\node[font=\scriptsize, tsBase, anchor=south east] at (axis cs:61,95) {0.8};
\end{axis}
\end{tikzpicture}
\caption{Memory--recomputation trade-off on Qwen3-8B (8 GPUs) for eager \sys and the \texttt{torch.compile} memory budget. Each method is normalized to its own full-checkpointing (0) and no-checkpointing (100) endpoints on both axes; labels indicate $f$ or the compiler budget.}
\label{fig:compile-q3}
\end{figure}
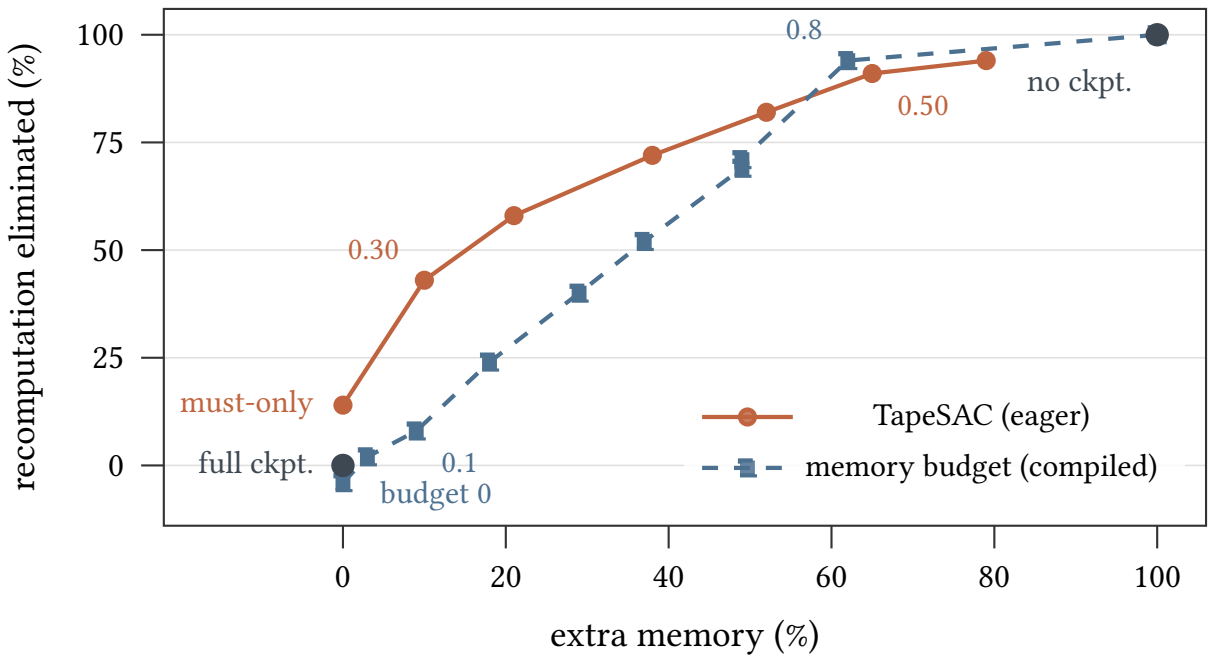

\subsection{Ablation Study}
\label{sec:eval-ablation}

We remove one design element at a time: the registration that makes
custom kernels replayable (\S\ref{sec:visibility}) and the two solver
options (\S\ref{sec:planning-solve} and~\S\ref{sec:planning-keys}).

\subsubsection{Kernel visibility}
\label{sec:eval-visibility}
Without registration, a custom kernel's outputs cannot be replayed
and are force-saved. The cost depends on the family. On the 27B model
at $f = 0.90$, leaving the GDN core unregistered force-saves its outputs in
every GDN block, so that the must-only peak alone exceeds the planner's
capacity; the planner saves nothing else, and the step is 8.8\% slower
than with the kernel registered. Leaving the root-mean-square (RMS)
normalizations unregistered force-saves both normalization outputs of every
layer, and the must-only sample step runs out of memory. On the 35B
model, the same two families slow the step by 5.2\% and 2.9\%. Leaving
FlashAttention-4 unregistered costs only 0.3\% on the 27B model, since its
outputs are among those the planner saves anyway, and leaving the
$qk$-normalization, RoPE and attention-gate kernels unregistered force-saves
nothing and costs 0.1\%. Registering these kernels is thus a
prerequisite for tensor-level checkpointing on these models rather than
an optimization: without it, the outputs of the kernels that dominate
activation memory can only be saved.

\subsubsection{Solver}
\label{sec:eval-solver}
The beam search is not what limits the plans: on the 27B model, a search
four times wider, keeping 96 sets per round with up to 16 candidates each,
predicts no shorter replay time at any of ten memory limits between
$f = 0.76$ and $0.94$. The two options, the bucket-edge charge and
per-block points, matter where memory is tight, each on a different
model (Table~\ref{tab:ablation}). On the 27B
model at $f = 0.78$, the FlashAttention-4 outputs of the 16
full-attention blocks need 7.4\% of device memory. Without the two
options, the room for saved values is 7.0\%, so no block saves them;
charging the must-only peak at the bucket edge raises the room to 8.3\%,
every block saves its output, and the step shortens by 0.55\,s, with the
bucket-edge charge alone giving the same. On the 35B model at
$f = 0.45$, the room barely changes, but per-block points let some vision
blocks give up a tensor that removes about half as much recomputation time per byte as
the FlashAttention-4 outputs, so that all 10 full-attention blocks can
save those outputs, and the step shortens by 87\,ms. At $f = 0.90$,
both options together still give the fastest step on both models, but by
at most 80\,ms: they matter when a valuable group does not fit on every
block of its region type, and otherwise mainly let the planner use the
memory it is given.

\subsection{Comparison with the Memory Budget of \texttt{torch.compile}}
\label{sec:eval-compile}

On the models we train, \texttt{torch.compile} is not an alternative: their
graph capture breaks before the partitioner runs (\S\ref{sec:bg-compile}),
and their hand-fused kernels leave the compiler's backend little to fuse,
which is its main source of speed. Where capture does succeed, the
limitations of \S\ref{sec:bg-compile} remain, and we measure their effect
on a model that the compiler can capture.

We compile torchtitan's Qwen3-8B (text only, 36 layers) one block at a
time and train it on 8 GPUs with FSDP and two sequences of 8192 tokens
per GPU. The eager side runs \sys from must-only to $f = 0.55$, the
compiled side the budget from 0 to 1, and each side also runs full
checkpointing and no checkpointing. Because compilation changes the step
time of every configuration, full checkpointing included,
Figure~\ref{fig:compile-q3} compares each side against its own endpoints:
extra memory is the memory used beyond full checkpointing, and
recomputation eliminated is the reduction in step time, each as a share of what
disabling checkpointing uses or removes.

With little extra memory, \sys eliminates several times more
recomputation than the budget: 43\% with 10\% of the extra memory, against
8\% with 9\% at budget 0.1. With none, must-only already eliminates 14\%,
whereas budget 0 is slower than full checkpointing ($-4\%$). The two
converge only with more than 60\% of the extra memory (91\% for \sys at
$f = 0.50$, 94\% for budget 0.8), so the budget is least effective when
memory is tight, which is when selective checkpointing is needed.

\begin{figure}[t]
\centering
\begin{tikzpicture}
\begin{axis}[
  width=\linewidth, height=5.4cm,
  xlabel={reserved memory, rank 0 (\% of device)}, ylabel={step time (s)},
  xmin=71.0, xmax=94.1, ymin=23.9, ymax=25.5,
  yticklabel style={/pgf/number format/fixed,/pgf/number format/precision=1},
  tick label style={font=\scriptsize}, label style={font=\footnotesize},
  legend style={font=\scriptsize, at={(0.98,0.97)}, anchor=north east,
    draw=none, fill=white, fill opacity=0.8, text opacity=1},
  legend cell align=left,
]
\addplot[tsOurs, thick, mark=*, mark size=1.6pt] coordinates
  {(73.3,24.877) (73.9,24.488) (76.3,24.309) (77.6,24.196) (79.3,24.097) (81.1,24.067) (87.4,24.146)};
\addlegendentry{\sys}
\addplot[only marks, mark=square*, mark size=2pt, tsFull] coordinates {(73.3,25.328)};
\addplot[only marks, mark=diamond*, mark size=2.6pt, tsBase] coordinates {(91.8,24.759)};
\addplot[only marks, mark=diamond, mark size=2.6pt, tsBase, thick] coordinates {(89.9,24.716)};
\node[font=\scriptsize, tsFull, anchor=west] at (axis cs:73.6,25.328) {full checkpointing};
\node[font=\scriptsize, tsOurs, anchor=west] at (axis cs:73.7,24.90) {must-only};
\node[font=\scriptsize, tsBase, anchor=south east] at (axis cs:92.3,24.79) {H1, SAC};
\node[font=\scriptsize, tsBase, anchor=north east] at (axis cs:90.4,24.70) {H1, replayed};
\node[font=\scriptsize, tsOurs, anchor=south west] at (axis cs:73.9,24.50) {0.84};
\node[font=\scriptsize, tsOurs, anchor=north] at (axis cs:79.3,24.08) {0.90};
\node[font=\scriptsize, tsOurs, anchor=south] at (axis cs:87.4,24.16) {0.99};
\end{axis}
\end{tikzpicture}
\caption{Step time versus rank-0 reserved memory for Qwen3.5-397B on 64 GPUs. Labels on the \sys curve indicate the memory limit $f$.}
\label{fig:frontier-397b}
\end{figure}
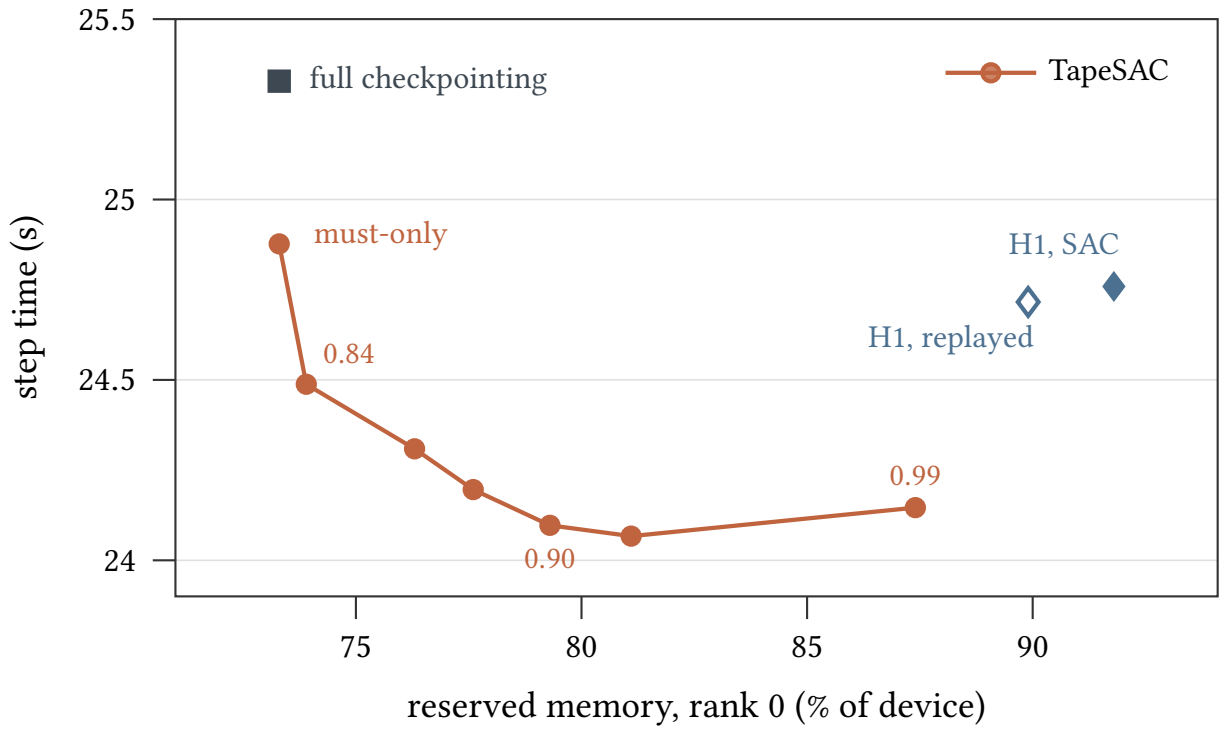

Controlled experiments on small models, compiled with Inductor on one
GPU, isolate two properties behind this. First, the budget bounds the
bytes saved for backward, not the peak. On a chain of 8 identical blocks compiled as one
graph, budget 0 saves no activations at all, yet the peak is unchanged
and the step is 38\% slower, whereas segment checkpointing, which
recomputes one segment at a time, lowers the peak by 26\%; at 16 and 32
blocks, the budget's peak is 1.59$\times$ and 1.76$\times$ that of
segment checkpointing at equal recomputation. Second, on top of
per-block checkpointing, which is how large models are trained, the
budget has almost no effect: on an 8-layer transformer, adding a budget
of 0.1 to 0.7 to full checkpointing leaves the step time unchanged and
lowers the peak by only 5\%.

\subsection{Scale and Real Data}
\label{sec:eval-397b}

\subsubsection{Qwen3.5-397B}
The results hold at scale, and \sys outperforms the hand-tuned policy H1
(Figure~\ref{fig:frontier-397b}). At the default $f = 0.90$, \sys
shortens the step by 4.9\% relative to full checkpointing; H1 shortens
it by only 2.2\% while raising reserved memory by 25\%, and \sys is
2.7\% faster than H1 with 13.6\% less memory on rank 0 (79.3\% against
91.8\% of device memory). This memory advantage is not limited to rank 0: although vision-token counts range from 230k to 334k across ranks, the maximum reserved memory across ranks exceeds rank 0's by only 1--3\%, because the shared plan is constrained by the rank with the least available memory.

\subsubsection{Real data}
\label{sec:eval-realdata}
On real data, token counts change from step to step; each new token
bucket triggers a solve, and token counts outside the measured range
trigger a new memory sample. We ran the 35B model with a local batch size of 16 for 500 steps, with full checkpointing and with \sys
at $f = 0.90$, in the same data order and on each of two pairs of nodes
(Table~\ref{tab:longrun}).

\begin{table}[t]
\centering
\footnotesize
\setlength{\tabcolsep}{4pt}
\caption{Real-data training results for Qwen3.6-35B-A3B over 500 steps. Full checkpointing and \sys ($f=0.90$) use the same data order on each of two node pairs; \sys ran first on node pair 1 and full checkpointing ran first on node pair 2.}
\label{tab:longrun}
\begin{tabular}{@{}lrr@{}}
\toprule
\textbf{Settings}                     & \textbf{Node pair 1} & \textbf{Node pair 2} \\
\midrule
Total time (min)                      &           &           \\
\quad full checkpointing              & 92.8      & 92.4      \\
\quad \sys                            & 85.4      & 82.1      \\
\quad change                          & $-8.0\%$  & $-11.1\%$ \\
Throughput change (global tokens/s)   & $+8.7\%$  & $+12.5\%$ \\
Step-time change                      &           &           \\
\quad steps without a solve or sample & $-10.7\%$ & $-12.6\%$ \\
\quad steps with a solve or sample (16) & $+8.3$\,s each & $+3.7$\,s each \\
\quad steps from 139 on               & $-12.3\%$ & $-13.0\%$ \\
Reserved memory (\%)                  &           &           \\
\quad full checkpointing              & 37.2      & 37.2      \\
\quad \sys                            & 75.5      & 75.7      \\
\bottomrule
\end{tabular}

\end{table}

Over 500 steps, \sys shortens training by 8.0\% and 11.1\% on the two
pairs of nodes, raising throughput by 8.7\% and 12.5\%. The number of
token buckets is bounded: after step 138, every bucket the data produces
already has a plan, and from there on \sys is 12.3\% and 13.0\% faster.
The difference comes from the 16 steps that solve a plan or sample
memory, all within the first 140 steps.

\subsection{Overhead and Generality}
\label{sec:eval-overhead}
\label{sec:eval-generality}

Recording and replay cost host time, not GPU time: about 24\,$\mu$s per
recorded operator, which GPU work hides from 2048 tokens per block
upward, far below the token counts we train with. Running the custom
kernels as registered custom operators costs 0.3--0.4\% of step time. The same \sys package runs all three dense and MoE models on 16--64 GPUs, requiring only a small model-specific adapter (\S\ref{sec:overview}).

\section{Conclusion}
\label{sec:conclusion}

\sys turns activation checkpointing from a policy that a person writes into
a problem that the system solves. A hand-tuned policy fixes which tensors
to keep, so its memory moves with every change of configuration; \sys
instead fixes the memory and solves for the tensors. Replaying a tape
recorded from each block's forward pass, it recomputes only what the
backward pass consumes, and its planner prices candidates with measured
kernel times and constrains the step's peak. It runs in eager PyTorch
without graph capture or edits to the model's forward code, and reproduces
the recomputed values bitwise.

On Qwen3.8-27B and Qwen3.6-35B-A3B, a single limit of 90\% of device memory
shortens the step by 6.3\% and 13.9\% relative to full checkpointing; at the
same or lower memory, \sys is up to 5.1\% faster than the hand-tuned
policies we evaluate, and it keeps its gain where they run out of memory.
On Qwen3.5-397B, it is 2.7\% faster than the policy hand-tuned for that
model with 14\% less memory, and over 500 steps on real data it shortens
training by 8--11\%.

The tape already tracks the size and lifetime of every value it records,
so the planner could decide more than what to save. Recording every
write the training step makes to device memory would let it also decide
where each tensor is placed, in device memory or offloaded to the host,
bringing allocator fragmentation into the plan; planning each rank
separately, so that slow ranks recompute less, may also mitigate
stragglers.

\bibliographystyle{ACM-Reference-Format}
\bibliography{refs}


\begin{thebibliography}{28}


\ifx \showCODEN    \undefined \def \showCODEN     #1{\unskip}     \fi
\ifx \showISBNx    \undefined \def \showISBNx     #1{\unskip}     \fi
\ifx \showISBNxiii \undefined \def \showISBNxiii  #1{\unskip}     \fi
\ifx \showISSN     \undefined \def \showISSN      #1{\unskip}     \fi
\ifx \showLCCN     \undefined \def \showLCCN      #1{\unskip}     \fi
\ifx \shownote     \undefined \def \shownote      #1{#1}          \fi
\ifx \showarticletitle \undefined \def \showarticletitle #1{#1}   \fi
\ifx \showURL      \undefined \def \showURL       {\relax}        \fi
\providecommand\bibfield[2]{#2}
\providecommand\bibinfo[2]{#2}
\providecommand\natexlab[1]{#1}
\providecommand\showeprint[2][]{arXiv:#2}
\makeatletter
\@ifundefined{NAT@parse@date}{}{\let\NAT@parse@date@orig\NAT@parse@date}
\@ifundefined{NAT@parse@date}{}{\def\NAT@parse@date#1#2#3#4#5#6@@{\NAT@parse@date@orig#1#2#3#4#5#6@@\def\NAT@tempyear{0000}\def\NAT@tempexlab{{?}}\ifx\NAT@year\NAT@tempyear\ifx\NAT@exlab\NAT@tempexlab\def\NAT@date{[n.\,d.]}\else\edef\NAT@date{[n.\,d.]\NAT@exlab}\fi\fi}}
\makeatother

\bibitem[Ansel et~al\mbox{.}(2024)]%
        {pytorch2}
\bibfield{author}{\bibinfo{person}{Jason Ansel}, \bibinfo{person}{Edward Yang},
  \bibinfo{person}{Horace He}, {et~al\mbox{.}}}
  \bibinfo{year}{2024}\natexlab{}.
\newblock \showarticletitle{{PyTorch} 2: Faster Machine Learning Through
  Dynamic {Python} Bytecode Transformation and Graph Compilation}. In
  \bibinfo{booktitle}{\emph{Proceedings of the 29th ACM International
  Conference on Architectural Support for Programming Languages and Operating
  Systems (ASPLOS)}}.
\newblock


\bibitem[Bartan et~al\mbox{.}(2023)]%
        {moccasin}
\bibfield{author}{\bibinfo{person}{Burak Bartan}, \bibinfo{person}{Haoming Li},
  \bibinfo{person}{Harris Teague}, \bibinfo{person}{Christopher Lott}, {and}
  \bibinfo{person}{Bistra Dilkina}.} \bibinfo{year}{2023}\natexlab{}.
\newblock \showarticletitle{Moccasin: Efficient Tensor Rematerialization for
  Neural Networks}. In \bibinfo{booktitle}{\emph{International Conference on
  Machine Learning (ICML)}}.
\newblock


\bibitem[Beaumont et~al\mbox{.}(2024)]%
        {rotor}
\bibfield{author}{\bibinfo{person}{Olivier Beaumont}, \bibinfo{person}{Lionel
  Eyraud-Dubois}, \bibinfo{person}{Julien Herrmann}, \bibinfo{person}{Alexis
  Joly}, {and} \bibinfo{person}{Alena Shilova}.}
  \bibinfo{year}{2024}\natexlab{}.
\newblock \showarticletitle{Optimal Re-Materialization Strategies for
  Heterogeneous Chains: How to Train Deep Neural Networks with Limited Memory}.
\newblock \bibinfo{journal}{\emph{ACM Trans. Math. Software}}
  \bibinfo{volume}{50}, \bibinfo{number}{2} (\bibinfo{year}{2024}),
  \bibinfo{pages}{1--38}.
\newblock
\href{https://doi.org/10.1145/3648633}{doi:\nolinkurl{10.1145/3648633}}


\bibitem[Chen et~al\mbox{.}(2016)]%
        {chen2016}
\bibfield{author}{\bibinfo{person}{Tianqi Chen}, \bibinfo{person}{Bing Xu},
  \bibinfo{person}{Chiyuan Zhang}, {and} \bibinfo{person}{Carlos Guestrin}.}
  \bibinfo{year}{2016}\natexlab{}.
\newblock \showarticletitle{Training Deep Nets with Sublinear Memory Cost}.
\newblock \bibinfo{journal}{\emph{arXiv preprint arXiv:1604.06174}}
  (\bibinfo{year}{2016}).
\newblock


\bibitem[Griewank and Walther(2000)]%
        {revolve}
\bibfield{author}{\bibinfo{person}{Andreas Griewank} {and}
  \bibinfo{person}{Andrea Walther}.} \bibinfo{year}{2000}\natexlab{}.
\newblock \showarticletitle{Algorithm 799: revolve: An Implementation of
  Checkpointing for the Reverse or Adjoint Mode of Computational
  Differentiation}.
\newblock \bibinfo{journal}{\emph{ACM Trans. Math. Software}}
  \bibinfo{volume}{26}, \bibinfo{number}{1} (\bibinfo{year}{2000}),
  \bibinfo{pages}{19--45}.
\newblock


\bibitem[Gusak et~al\mbox{.}(2025)]%
        {hiremate}
\bibfield{author}{\bibinfo{person}{Julia Gusak}, \bibinfo{person}{Xunyi Zhao},
  \bibinfo{person}{Th{\'e}otime Le~Hellard}, \bibinfo{person}{Zhe Li},
  \bibinfo{person}{Lionel Eyraud-Dubois}, {and} \bibinfo{person}{Olivier
  Beaumont}.} \bibinfo{year}{2025}\natexlab{}.
\newblock \showarticletitle{{HiRemate}: Hierarchical Approach for Efficient
  Re-materialization of Neural Networks}. In
  \bibinfo{booktitle}{\emph{Proceedings of the 42nd International Conference on
  Machine Learning (ICML)}} \emph{(\bibinfo{series}{Proceedings of Machine
  Learning Research}, Vol.~\bibinfo{volume}{267})}.
  \bibinfo{pages}{21418--21443}.
\newblock


\bibitem[He and Yu(2023)]%
        {fusionaware}
\bibfield{author}{\bibinfo{person}{Horace He} {and} \bibinfo{person}{Shangdi
  Yu}.} \bibinfo{year}{2023}\natexlab{}.
\newblock \showarticletitle{Transcending Runtime-Memory Tradeoffs in
  Checkpointing by Being Fusion Aware}. In
  \bibinfo{booktitle}{\emph{Proceedings of Machine Learning and Systems
  (MLSys)}}.
\newblock


\bibitem[Jain et~al\mbox{.}(2020)]%
        {checkmate}
\bibfield{author}{\bibinfo{person}{Paras Jain}, \bibinfo{person}{Ajay Jain},
  \bibinfo{person}{Aniruddha Nrusimha}, \bibinfo{person}{Amir Gholami},
  \bibinfo{person}{Pieter Abbeel}, \bibinfo{person}{Joseph Gonzalez},
  \bibinfo{person}{Kurt Keutzer}, {and} \bibinfo{person}{Ion Stoica}.}
  \bibinfo{year}{2020}\natexlab{}.
\newblock \showarticletitle{Checkmate: Breaking the Memory Wall with Optimal
  Tensor Rematerialization}. In \bibinfo{booktitle}{\emph{Proceedings of
  Machine Learning and Systems (MLSys)}}.
\newblock


\bibitem[Kirisame et~al\mbox{.}(2021)]%
        {dtr}
\bibfield{author}{\bibinfo{person}{Marisa Kirisame}, \bibinfo{person}{Steven
  Lyubomirsky}, \bibinfo{person}{Altan Haan}, \bibinfo{person}{Jennifer
  Brennan}, \bibinfo{person}{Mike He}, \bibinfo{person}{Jared Roesch},
  \bibinfo{person}{Tianqi Chen}, {and} \bibinfo{person}{Zachary Tatlock}.}
  \bibinfo{year}{2021}\natexlab{}.
\newblock \showarticletitle{Dynamic Tensor Rematerialization}. In
  \bibinfo{booktitle}{\emph{International Conference on Learning
  Representations (ICLR)}}.
\newblock


\bibitem[Korthikanti et~al\mbox{.}(2023)]%
        {megatron-selective}
\bibfield{author}{\bibinfo{person}{Vijay Korthikanti}, \bibinfo{person}{Jared
  Casper}, \bibinfo{person}{Sangkug Lym}, \bibinfo{person}{Lawrence McAfee},
  \bibinfo{person}{Michael Andersch}, \bibinfo{person}{Mohammad Shoeybi}, {and}
  \bibinfo{person}{Bryan Catanzaro}.} \bibinfo{year}{2023}\natexlab{}.
\newblock \showarticletitle{Reducing Activation Recomputation in Large
  Transformer Models}. In \bibinfo{booktitle}{\emph{Proceedings of Machine
  Learning and Systems (MLSys)}}.
\newblock


\bibitem[Lepikhin et~al\mbox{.}(2021)]%
        {lepikhin2021gshard}
\bibfield{author}{\bibinfo{person}{Dmitry Lepikhin},
  \bibinfo{person}{HyoukJoong Lee}, \bibinfo{person}{Yuanzhong Xu},
  \bibinfo{person}{Dehao Chen}, \bibinfo{person}{Orhan Firat},
  \bibinfo{person}{Yanping Huang}, \bibinfo{person}{Maxim Krikun},
  \bibinfo{person}{Noam Shazeer}, {and} \bibinfo{person}{Zhifeng Chen}.}
  \bibinfo{year}{2021}\natexlab{}.
\newblock \showarticletitle{{GShard}: Scaling Giant Models with Conditional
  Computation and Automatic Sharding}. In
  \bibinfo{booktitle}{\emph{International Conference on Learning
  Representations (ICLR)}}.
\newblock


\bibitem[Li et~al\mbox{.}(2023)]%
        {li2023sequence}
\bibfield{author}{\bibinfo{person}{Shenggui Li}, \bibinfo{person}{Fuzhao Xue},
  \bibinfo{person}{Chaitanya Baranwal}, \bibinfo{person}{Yongbin Li}, {and}
  \bibinfo{person}{Yang You}.} \bibinfo{year}{2023}\natexlab{}.
\newblock \showarticletitle{Sequence Parallelism: Long Sequence Training from
  System Perspective}. In \bibinfo{booktitle}{\emph{Proceedings of the 61st
  Annual Meeting of the Association for Computational Linguistics (ACL)}}.
  \bibinfo{pages}{2391--2404}.
\newblock
\href{https://doi.org/10.18653/v1/2023.acl-long.134}{doi:\nolinkurl{10.18653/v1/2023.acl-long.134}}


\bibitem[Liao et~al\mbox{.}(2023)]%
        {liao2023mimose}
\bibfield{author}{\bibinfo{person}{Jianjin Liao}, \bibinfo{person}{Mingzhen
  Li}, \bibinfo{person}{Hailong Yang}, \bibinfo{person}{Qingxiao Sun},
  \bibinfo{person}{Biao Sun}, \bibinfo{person}{Jiwei Hao},
  \bibinfo{person}{Tianyu Feng}, \bibinfo{person}{Fengwei Yu},
  \bibinfo{person}{Shengdong Chen}, \bibinfo{person}{Ye Tao},
  \bibinfo{person}{Zicheng Zhang}, \bibinfo{person}{Zhongzhi Luan}, {and}
  \bibinfo{person}{Depei Qian}.} \bibinfo{year}{2023}\natexlab{}.
\newblock \showarticletitle{Exploiting Input Tensor Dynamics in Activation
  Checkpointing for Efficient Training on {GPU}}. In
  \bibinfo{booktitle}{\emph{IEEE International Parallel and Distributed
  Processing Symposium (IPDPS)}}. \bibinfo{pages}{156--166}.
\newblock
\href{https://doi.org/10.1109/IPDPS54959.2023.00025}{doi:\nolinkurl{10.1109/IPDPS54959.2023.00025}}


\bibitem[Liu et~al\mbox{.}(2024)]%
        {liu2024ringattention}
\bibfield{author}{\bibinfo{person}{Hao Liu}, \bibinfo{person}{Matei Zaharia},
  {and} \bibinfo{person}{Pieter Abbeel}.} \bibinfo{year}{2024}\natexlab{}.
\newblock \showarticletitle{{RingAttention} with Blockwise Transformers for
  Near-Infinite Context}. In \bibinfo{booktitle}{\emph{International Conference
  on Learning Representations (ICLR)}}.
\newblock


\bibitem[{Meta Platforms, Inc.}(2026)]%
        {torch-remat}
\bibfield{author}{\bibinfo{person}{{Meta Platforms, Inc.}}}
  \bibinfo{year}{2026}\natexlab{}.
\newblock \bibinfo{title}{torch\_remat: Explicit Activation Rematerialization
  Helpers for {PyTorch}}.
\newblock \bibinfo{howpublished}{\url{https://github.com/meta-pytorch/remat}}.
\newblock
\newblock
\shownote{Version 0.2.0}.


\bibitem[Peng et~al\mbox{.}(2020)]%
        {capuchin}
\bibfield{author}{\bibinfo{person}{Xuan Peng}, \bibinfo{person}{Xuanhua Shi},
  \bibinfo{person}{Hulin Dai}, \bibinfo{person}{Hai Jin},
  \bibinfo{person}{Weiliang Ma}, \bibinfo{person}{Qian Xiong},
  \bibinfo{person}{Fan Yang}, {and} \bibinfo{person}{Xuehai Qian}.}
  \bibinfo{year}{2020}\natexlab{}.
\newblock \showarticletitle{Capuchin: Tensor-based {GPU} Memory Management for
  Deep Learning}. In \bibinfo{booktitle}{\emph{Proceedings of the 25th ACM
  International Conference on Architectural Support for Programming Languages
  and Operating Systems (ASPLOS)}}.
\newblock


\bibitem[{PyTorch Contributors}(2026a)]%
        {pytorch-sac-ilp}
\bibfield{author}{\bibinfo{person}{{PyTorch Contributors}}.}
  \bibinfo{year}{2026}\natexlab{a}.
\newblock \bibinfo{title}{\texttt{torch/distributed/\_tools/sac\_ilp.py},
  {PyTorch} 2.13}.
\newblock
  \bibinfo{howpublished}{\url{https://github.com/pytorch/pytorch/blob/v2.13.0/torch/distributed/_tools/sac_ilp.py}}.
\newblock


\bibitem[{PyTorch Contributors}(2026b)]%
        {pytorch-partitioner}
\bibfield{author}{\bibinfo{person}{{PyTorch Contributors}}.}
  \bibinfo{year}{2026}\natexlab{b}.
\newblock \bibinfo{title}{\texttt{torch/\_functorch/partitioners.py}, {PyTorch}
  2.13}.
\newblock
  \bibinfo{howpublished}{\url{https://github.com/pytorch/pytorch/blob/v2.13.0/torch/_functorch/partitioners.py}}.
\newblock


\bibitem[Rajbhandari et~al\mbox{.}(2020)]%
        {zero}
\bibfield{author}{\bibinfo{person}{Samyam Rajbhandari}, \bibinfo{person}{Jeff
  Rasley}, \bibinfo{person}{Olatunji Ruwase}, {and} \bibinfo{person}{Yuxiong
  He}.} \bibinfo{year}{2020}\natexlab{}.
\newblock \showarticletitle{{ZeRO}: Memory Optimizations Toward Training
  Trillion Parameter Models}. In \bibinfo{booktitle}{\emph{Proceedings of the
  International Conference for High Performance Computing, Networking, Storage
  and Analysis (SC)}}.
\newblock


\bibitem[Schuler et~al\mbox{.}(2022)]%
        {schuler2022xengine}
\bibfield{author}{\bibinfo{person}{Manuela Schuler}, \bibinfo{person}{Richard
  Membarth}, {and} \bibinfo{person}{Philipp Slusallek}.}
  \bibinfo{year}{2022}\natexlab{}.
\newblock \showarticletitle{{XEngine}: Optimal Tensor Rematerialization for
  Neural Networks in Heterogeneous Environments}.
\newblock \bibinfo{journal}{\emph{ACM Transactions on Architecture and Code
  Optimization}} \bibinfo{volume}{20}, \bibinfo{number}{1}
  (\bibinfo{year}{2022}), \bibinfo{pages}{17:1--17:25}.
\newblock
\href{https://doi.org/10.1145/3568956}{doi:\nolinkurl{10.1145/3568956}}


\bibitem[Shoeybi et~al\mbox{.}(2019)]%
        {shoeybi2019megatron}
\bibfield{author}{\bibinfo{person}{Mohammad Shoeybi}, \bibinfo{person}{Mostofa
  Patwary}, \bibinfo{person}{Raul Puri}, \bibinfo{person}{Patrick LeGresley},
  \bibinfo{person}{Jared Casper}, {and} \bibinfo{person}{Bryan Catanzaro}.}
  \bibinfo{year}{2019}\natexlab{}.
\newblock \showarticletitle{{Megatron-LM}: Training Multi-Billion Parameter
  Language Models Using Model Parallelism}.
\newblock \bibinfo{journal}{\emph{arXiv preprint arXiv:1909.08053}}
  (\bibinfo{year}{2019}).
\newblock


\bibitem[Steiner et~al\mbox{.}(2023)]%
        {model}
\bibfield{author}{\bibinfo{person}{Benoit Steiner}, \bibinfo{person}{Mostafa
  Elhoushi}, \bibinfo{person}{Jacob Kahn}, {and} \bibinfo{person}{James
  Hegarty}.} \bibinfo{year}{2023}\natexlab{}.
\newblock \showarticletitle{{MODeL}: Memory Optimizations for Deep Learning}.
  In \bibinfo{booktitle}{\emph{International Conference on Machine Learning
  (ICML)}}.
\newblock


\bibitem[Yang et~al\mbox{.}(2025)]%
        {gdn}
\bibfield{author}{\bibinfo{person}{Songlin Yang}, \bibinfo{person}{Jan Kautz},
  {and} \bibinfo{person}{Ali Hatamizadeh}.} \bibinfo{year}{2025}\natexlab{}.
\newblock \showarticletitle{Gated Delta Networks: Improving {Mamba2} with Delta
  Rule}. In \bibinfo{booktitle}{\emph{International Conference on Learning
  Representations (ICLR)}}.
\newblock


\bibitem[Yuan et~al\mbox{.}(2024)]%
        {yuan2024activation}
\bibfield{author}{\bibinfo{person}{Tailing Yuan}, \bibinfo{person}{Yuliang
  Liu}, \bibinfo{person}{Xucheng Ye}, \bibinfo{person}{Shenglong Zhang},
  \bibinfo{person}{Jianchao Tan}, \bibinfo{person}{Bin Chen},
  \bibinfo{person}{Chengru Song}, {and} \bibinfo{person}{Di Zhang}.}
  \bibinfo{year}{2024}\natexlab{}.
\newblock \showarticletitle{Accelerating the Training of Large Language Models
  using Efficient Activation Rematerialization and Optimal Hybrid Parallelism}.
  In \bibinfo{booktitle}{\emph{USENIX Annual Technical Conference (USENIX
  ATC)}}. \bibinfo{pages}{545--561}.
\newblock


\bibitem[Zhang et~al\mbox{.}(2023)]%
        {coop}
\bibfield{author}{\bibinfo{person}{Jianhao Zhang}, \bibinfo{person}{Shihan Ma},
  \bibinfo{person}{Peihong Liu}, {and} \bibinfo{person}{Jinhui Yuan}.}
  \bibinfo{year}{2023}\natexlab{}.
\newblock \showarticletitle{Coop: Memory is not a Commodity}. In
  \bibinfo{booktitle}{\emph{Advances in Neural Information Processing Systems
  (NeurIPS)}}.
\newblock


\bibitem[Zhao et~al\mbox{.}(2023b)]%
        {rockmate}
\bibfield{author}{\bibinfo{person}{Xunyi Zhao}, \bibinfo{person}{Th{\'e}otime
  Le~Hellard}, \bibinfo{person}{Lionel Eyraud-Dubois}, \bibinfo{person}{Julia
  Gusak}, {and} \bibinfo{person}{Olivier Beaumont}.}
  \bibinfo{year}{2023}\natexlab{b}.
\newblock \showarticletitle{Rockmate: an Efficient, Fast, Automatic and Generic
  Tool for Re-materialization in {PyTorch}}. In
  \bibinfo{booktitle}{\emph{International Conference on Machine Learning
  (ICML)}}.
\newblock


\bibitem[Zhao et~al\mbox{.}(2023a)]%
        {fsdp}
\bibfield{author}{\bibinfo{person}{Yanli Zhao}, \bibinfo{person}{Andrew Gu},
  \bibinfo{person}{Rohan Varma}, \bibinfo{person}{Liang Luo},
  \bibinfo{person}{Chien-Chin Huang}, \bibinfo{person}{Min Xu},
  \bibinfo{person}{Less Wright}, \bibinfo{person}{Hamid Shojanazeri},
  \bibinfo{person}{Myle Ott}, \bibinfo{person}{Sam Shleifer},
  \bibinfo{person}{Alban Desmaison}, \bibinfo{person}{Can Balioglu},
  \bibinfo{person}{Pritam Damania}, \bibinfo{person}{Bernard Nguyen},
  \bibinfo{person}{Geeta Chauhan}, \bibinfo{person}{Yuchen Hao},
  \bibinfo{person}{Ajit Mathews}, {and} \bibinfo{person}{Shen Li}.}
  \bibinfo{year}{2023}\natexlab{a}.
\newblock \showarticletitle{{PyTorch} {FSDP}: Experiences on Scaling Fully
  Sharded Data Parallel}.
\newblock \bibinfo{journal}{\emph{Proceedings of the VLDB Endowment}}
  \bibinfo{volume}{16}, \bibinfo{number}{12} (\bibinfo{year}{2023}),
  \bibinfo{pages}{3848--3860}.
\newblock


\bibitem[Zhu et~al\mbox{.}(2025)]%
        {mist}
\bibfield{author}{\bibinfo{person}{Zhanda Zhu}, \bibinfo{person}{Christina
  Giannoula}, \bibinfo{person}{Muralidhar Andoorveedu}, \bibinfo{person}{Qidong
  Su}, \bibinfo{person}{Karttikeya Mangalam}, \bibinfo{person}{Bojian Zheng},
  {and} \bibinfo{person}{Gennady Pekhimenko}.} \bibinfo{year}{2025}\natexlab{}.
\newblock \showarticletitle{Mist: Efficient Distributed Training of Large
  Language Models via Memory-Parallelism Co-Optimization}. In
  \bibinfo{booktitle}{\emph{Proceedings of the Twentieth European Conference on
  Computer Systems (EuroSys)}}.
\newblock
\href{https://doi.org/10.1145/3689031.3717461}{doi:\nolinkurl{10.1145/3689031.3717461}}


\end{thebibliography}

\end{document}